\documentclass[preprint,showpacs,preprintnumbers,amsmath,amssymb]{revtex4}
\usepackage{graphicx,epsfig,dcolumn,bm,epic,eepic,float}%
\usepackage{amsmath}
\usepackage{latexsym}
\usepackage{color}
\usepackage{cancel}
\usepackage{makeidx,shortvrb,latexsym}
\usepackage{slashed}
\begin{document}
\unitlength 1 cm
\newcommand{\nn}{\nonumber}
\newcommand{\vk}{\vec k}
\newcommand{\vp}{\vec p}
\newcommand{\vq}{\vec q}
\newcommand{\vkp}{\vec {k'}}
\newcommand{\vpp}{\vec {p'}}
\newcommand{\vqp}{\vec {q'}}
\newcommand{\bk}{{\bf k}}
\newcommand{\bp}{{\bf p}}
\newcommand{\bq}{{\bf q}}
\newcommand{\br}{{\bf r}}
\newcommand{\bR}{{\bf R}}
\newcommand{\up}{\uparrow}
\newcommand{\down}{\downarrow}
\newcommand{\fns}{\footnotesize}
\newcommand{\ns}{\normalsize}
\newcommand{\cdag}{c^{\dagger}}
\title {The P-P(\={P}) cross-section of   isolated single-photon production in the    $k_t$-factorization and
different angular ordering   unintegrated parton distributions
frameworks}
\author{$R. \; Aminzadeh\;Nik^1$ }
\author{$M. \; Modarres^1$ }
\altaffiliation {Corresponding author, Email: mmodares@ut.ac.ir,
Tel:+98-21-61118645, Fax:+98-21-88004781.}
\author{$N.\; Olanj^2$}
\author{$R.\; Taghavi^1$}
\affiliation {$^1$Department of Physics, University of $Tehran$,
1439955961, $Tehran$, Iran.} \affiliation {$^2$Physics Department,
Faculty of Science, $Bu$-$Ali\; Sina$ University, 65178, $Hamedan$,
Iran.}
\begin{abstract}
In the present work, it is  intended to calculate and  study the
 single and double differential cross-sections
  of the prompt single-photon production    as a function
of   produced single-photon transverse momentum and rapidity,   in
the high-energy P-P(\={P})  colliders, such as LHC and TEVATRON. The
 differential cross-sections of prompt single-photon production   are calculated in
 the $k_t$-factorization frameworks using   various angular ordering
unintegrated parton distribution functions (UPDF), namely the Kimber
et al. and Martin et al. procedures. These scheme-dependent UPDF are
generated in the leading and   next-to-leading order levels   to
predict and analyze the different partonic contributions to the
above cross-sections. The above two procedures utilize the
phenomenological parton distribution functions (PDF) libraries of
Martin et al., i.e.,  MMHT2014. It is shown that the calculated
prompt single-photon production  differential cross-sections in the
above frameworks are relatively successful  in generating
satisfactory results   compared to the experimental data of
different collaborations, i.e., CDF (2017), ATLAS (2017), CMS (2011)
and D0 (2006), as well as the other theoretical predictions such as
collinear factorization Monte Carlo calculations (the JETPHOX, SHERPA,
PYTHIA, and MCFM methods). Also, for a closer precision, the
differential cross-section for the NLO gluon-gluon  , quark-gluon  and
 quark-(anti)quark   processes  are calculated. An
extensive discussion and comparison are made regarding (i) the
behavior of the contributing partonic sub-processes, (ii) the
possible double-counting between the $2\to 2$ and $2\to 3$
sub-processes, i.e., gluons-fusion, in our calculated prompt
single-photon production differential cross-sections and (iii) the
sensibility check of our results to the different angular ordering
constraints.
\end{abstract}

\pacs{12.38.Bx, 13.85.Qk, 13.60.-r
\\ \textbf{Keywords:} unintegrated parton distribution functions,
single-photon production, $k_t$-factorization, CMS, ATLAS, D0 and
CDF collaborations, pQCD, collinear factorization.} \maketitle
\section{Introduction}
\label{sec:intro} The measurement of  prompt   single-photon
production   differential
  cross-sections   in the proton-(anti)proton collisions is a
powerful and vital test for a scrutiny check of   theoretical pQCD.
The results of these measurements are used to: (1) determine and
study the QCD strong coupling constant, $\alpha_s$,
\cite{Albino,Klasen,K1,K2}, (2) investigate the resummation of
threshold logarithmic behavior in the pQCD and   electroweak
corrections \cite{Schwartz} and, (3) to analyze     the background
processes for the measurements of the Higgs boson \cite{Atlas2016},
as well as the study of   photon isolations and   fragmentation
behaviors \cite{D02000,Farrar}.

The dominant prompt single-photon production in the hadron-hadron
collisions at the CERN (LHC) and Fermilab (TEVATRON) laboratories
proceeds via the $q(\bar q )+g \rightarrow \gamma+q(\bar q )$
Compton scattering process \cite{Atlas2017}  in the large photon
transverse momentum. But this is not the case, at the small momentum
regions \cite{au}. Therefore, these measurements are sensitive to
the gluon density of proton or anti-proton, in the leading order
(LO) \cite{7,8,9,10,11,12,13,14,15,16}. The results of these
reactions play an important role in determining the parton
distribution functions (PDF), i.e., $a_i(x,\mu^2)$, where $x$ and
$\mu^2$ are the longitudinal momentum fraction and the hard scale,
respectively. On the other hand, because of the above transverse
momentum dependent, they can also
   provide a useful piece of   information
about the unintegrated PDF (UPDF), $f_i(x,k_t^2,\mu^2)$, where $k_t$
is the parton transverse momentum  \cite{PRD75,amin}.

The present report  is the extension of our recent calculations for
the prompt photon pair production \cite{amin}. It is intended to
evaluate and analyze the behaviors of different UPDF and   angular
ordering constraints   in calculating the prompt single-photon
production differential cross-sections, i.e., $P+P(\bar P)
\rightarrow \gamma +X$.
 This process  mostly happens in the parton-(anti)parton high-energy scattering.
The analysis of the single and double differential cross-sections of
prompt single-photon production data are performed at the LHC and
TEVATRON colliders  by many experimental collaborations, at the some
center-of-mass energies. At LHC, the ATLAS and   CMS collaborations
measure the prompt single-photon production   events at $7$, $8$
and, $13$ TeV center of mass energies
\cite{Atlas2017,17,18,19,20,21,22,cms}. Similar measurements were
also performed at the TEVATRON at Fermilab by the D0 and CDF
collaborations at 1.96 TeV \cite{D02005,CDF2017,23,24}. However, we
intend to focus on the recent ones, such as the CDF and D0, CMS and
ATLAS collaborations at the center of mass energies of 1.96, 7 and,
13 TeV \cite{D02005,CDF2017,Atlas2017,cms}, respectively.

Today,   different  experimental collaborations conventionally use
the PDF   to describe and predict their  data  by performing the
collinear factorization simulation and the Monte Carlo techniques,
such as JETPHOX, SHERPA, PYTHIA, and MCFM
\cite{sherpa,pythia,jetphox,mcfm}. The SHERPA Monte Carlo event
generator provides the simulation of high energy reactions of
particles in the hadron-hadron collisions. The SHERPA results
include all of the tree-level matrix element amplitudes with the
one-photon and up to the three partons. This method features a
parton-jet matching procedure to avoid an overlap between the
phase-space descriptions given by the fixed-order matrix-element
sub-processes and the showering/ hadronization, in the multi-jet
simulation \cite{CDF2017,sherpa}. The JETPHOX technique is designed to calculate
the differential cross-section of hadron-hadron reactions to the
photon/hadron plus jet. By integrating over the jet, one can also
get the single inclusive photon-hadron cross-section at the
next-to-leading order (NLO) level \cite{D02005,jetphox}. The PYTHIA
program is a standard tool for
 the generation of events in the high-energy collisions and, its
predictions include the $2 \to 2$
  matrix-element of   sub-processes, in which the higher-order
collinear factorization corrections are included by the initial and
final-state parton showers \cite{CDF2017,pythia}. Finally, the MCFM
is a parton-level Monte Carlo computing code that gives the NLO
predictions by including the non-perturbative fragmentation at the
LO level for a range of processes at the hadron colliders
\cite{cms,mcfm}.

The above PDF are the solutions of the
Dokshitzer-Gribov-Lipatov-Altarelli-Parisi (DGLAP) evolution
equations \cite{DGLAP1,DGLAP2, DGLAP3, DGLAP4}, enriched by the
extended collinear factorization supplements like the parton
fragmentation  and the parton showers effects. The DGLAP evolution
equations, however, are based on the strong ordering assumption,
which generally neglects the transverse momentum, $k_t$, of the
emitted partons. It is frequently pointed out that undermining the
contributions coming from the transverse momentum of   partons may
severely harm the accuracy of   calculations. This indicates the
necessity of introducing some transverse momentum dependent (TMD)
PDF, e.g., through the Catani-Ciafaloni-Fiorani-Marchesini, (CCFM)
\cite{CCFM1,CCFM2,CCFM3,CCFM4,CCFM5,Q-CCFMI,Q-CCFMII}  or the Balitsky-Fadin-Kuraev-Lipatov (BFKL)
\cite{BFKL1,BFKL2, BFKL3, BFKL4, BFKL5} evolution equations. In
general,  using and solving the CCFM  and
  BFKL equations are
 difficult and proved to face complexity.  On the other hand,
 the main feature of the CCFM equation, i.e., angular ordering constraint, can be particularly used for the gluon
evolution.

One should note that, in the most of high energy hadron-hadron
collisions, the $x$ values of colliding partons are rather small
($x_1x_2\cong {\mu^2 \over s}\ll 1$, where $s$ is the center of mass
energy), since  the   fraction of the incoming longitudinal momentum
becomes large only when the very massive states near the threshold
are produced. So, besides collinear factorization calculations, the
$k_t$-factorization approach can be applied for re-summing the hard
cross-sections at the same order. The UPDF $k_t$-factorization
scheme is specially designed for the small x limit in which the hard
scale is fixed, and the energies are rather very high. However, there are
also various transverse momentum dependent factorization (TMDF)
\cite{amin,co}, which are usually used in the semi-inclusive
processes in the non-perturbative region \cite{amin}.

To overcome these problems, the Martin group developed the
Kimber-Martin-Ryskin (KMR) and the Martin-Ryskin-Watt (MRW)
approaches \cite{KMR,KMR1,kimber2,MRW,WATT} in the
$k_t$-factorization framework. Each approach   is built based on the
(LO and NLO)  DGLAP evolution equations and improved with different
visualizations of the angular ordering constraint. The KMR and MRW
formalisms in the  LO and   NLO levels were investigated intensely
in the recent years; see the references
\cite{Dijet,W/Z-LHCb,W/Z-NLO,FL-dipole,FL,Mod6,Mod5,Mod4,Mod3,Mod2,Mod1,amin,AMIN}.

The $k_t$-factorization becomes more accurate by using doubly UPDF
(DUPDF), i.e., ($z,k_t$)-factorization, in which the partons have
the virtuality $-{k_t^2\over (1-z)}$ \cite{WattWZ}. It is based on
the DDT formula \cite{DDT} in the pQCD with this difference that it
goes beyond the double-leading-logarithmic approximation (DLLA)
\cite{WattWZ,DDT}. We hope in our future works we could examine the
effect of DUPDF on different observables \cite{kord}. On the other hand, as it
is mentioned in the reference \cite{WattWZ}, the application of the
integrated PDF in the last evolution step, should be generated
through a new global fit to the experimental data using the
$k_t$-factorization procedure. In the reference \cite{WattWZ}, this
effect was estimated to lower the proton structure functions by 10
percent. However, we
 should make this note that, as it is discussed in the references
\cite{MRW,WattWZ,amin}, the $k_t$-factorization should not be as
good as collinear factorization approaches, on the other hand, it is
more simplistic in this scenes that it saves an enormous
computational time \cite{MRW,WattWZ,amin}.

As we said before,   we intend to study the  prompt single-photon
production and   calculate  the differential cross-sections
numerically   to investigate the different behaviors of the
cross-section by implementing various angular ordering constraints.
For this purpose, we use the lowest order  sets of matrix elements
and
 the $k_t$-factorization with the input  KMR \cite{KMR}, and
the MRW UPDF in the LO and NLO levels \cite{MRW}. Afterward, our
results are compared to the existing experimental data given by the
D0, CDF, ATLAS and, CMS collaborations
\cite{D02005,CDF2017,Atlas2017,cms}, as well as some of the
collinear factorization works based on the Monte Carlo simulations
\cite{jetphox,sherpa,pythia,mcfm} (JETPHOX, SHERPA, PYTHIA, and
MCFM).

There exist some articles in which the single-photon productions
were investigated  with the $k_t$-factorization formalism
\cite{KMR1,
kimber2,lips,lips0,lips1,lips5,lips7,lips8,lip2007,PRD75}, where
some conflicts acquire  in   their calculations. We present some
details about these open problems in the section \ref{sec:results}.
It is also noted in the reference \cite{PRD75} that the UPDF
generated    in some of the above references
\cite{lips,lips0,lips1,lips5,lips7,lips8,lip2007}, are not the
original version introduced by Kimber et al. \cite{KMR} nor the one
suggested by Martin et al \cite{MRW}. So these discrepancies should
be investigated in details, as it is pointed out in the references
\cite{PRD75,amin}. Also, a detailed study of the impact of the NLO
sub-processes i.e.  the gluon-gluon,  quark-gluon and quark-(anti)quark (see the figure 10) 
  are  performed and the results are presented for
comparison in the section III. It should be noted that this prompt
is calculated directly from the NLO perturbation of these processes.
However, there are other methods that can indirectly calculate the
share of these sub-processes. In these methods, the effects of
higher-order perturbation  are considered as an effective correction
in the LO order \cite{tagh1}.

Similar to the discussion made in  our photon pair report
\cite{amin}: (1) The off-shell matrix elements that are needed for
differential cross-sections calculations are gauge invariance,
especially since we are working in the small $x$ region (see, e.g., the references
\cite{amin,lips0} for details). (2) There is not any double-counting
between the $2\rightarrow 2$ and $2\rightarrow 3$ sub-processes
\cite{W/Z-LHCb}. Because  one should consider the KMR or MRW parton
densities in the $k_t$-factorization formalism correspond to the
probability functions similar to the PDF in the collinear case,
since all the splittings and the real emissions of the partons,
including the last emission, are factorized in the UPDF
\cite{KIMBER} (note that the DGLAP evolution equation derived by
integrating over the transverse momentum of partons by ignoring the
$k_t$ dependent of the PDF). On the other hand, any changes into the
UPDF certainly influence the normalization relation between the UPDF
and the original PDF (see the equation (\ref{eq0065})). However,
similar to the reference \cite{amin}, both points (1) and (2) will be discussed
through this paper.

In the following, first, the theoretical framework of  prompt
single-photon production events (\ref{cross}),  a brief introduction
to the KMR and
 LO  and  NLO MRW prescriptions (\ref{UPDF}) and, the experimental conditions (\ref{cond}) are presented in the
section \ref{sec:framework}. The section \ref{sec:results} is
devoted to our numerical results ({\ref{num})   and discussions
(\ref{dis}). Finally, a brief conclusion is presented in the section
\ref{sec:conc}.
\section{The Theoretical framework of differential cross-sections prompt single-photon production}
\label{sec:framework}
\subsection{The cross-section}
\label{cross} The   prompt single-photon production events are an
important and useful process to study  the PDF. Therefore, it would
be interesting  to investigate the partonic structure of the proton
at each energy while applying the necessary constraints on the
experiments. According to the reference \cite{D02005}  the photons
generated in the hard interaction between two partons  mainly come
from the quark-gluon Compton scattering or the quark-anti-quark
annihilation in the hadron collisions. Generally, the photons are
so-called prompt, if they are coupled to the interacting quarks
\cite{8,a2,a3,a4}. On the other hand, these photons do not produce
from the meson decays. Such events are mostly happened  in the
collinear factorization directly through the $q(\bar q)+g
\longrightarrow\gamma+ q (\bar q)$ and $q+\bar q\longrightarrow
\gamma+ g$ processes \cite{D02013}. Usually, in the experimental
measurements and theoretical calculations, it is focused on those
processes  which provide a direct probe of the hard sub-process
dynamics, since the produced photons are mostly insensitive to the
final-state hadronization
\cite{D02000,26,27,28,29,D02005,31,32,33,33p}. At the
 LO collinear
factorization and  in the prompt single-photon production where the
photon transverse momentum  is rather large, the quark-gluon Compton
scattering has a larger contribution with respect to the
quark-antiquark annihilation process. However, this calculation was
done up to the NLO
 collinear
factorization \cite{nlosp1,jetphox,nlosp3,nlosp4} and the results are
in agreement to the data, but some open questions are remaining
\cite{lip2008,re1,re2}.

In general, the   prompt single-photon production in the
hadron-hadron colliders can be described as follows:
$$
A+B\to a+b\to \gamma+X,
$$
where $a$ and $b$ are the incoming partons which are emitted by the
parent hadrons (A and B). In this work, the sets of LO $2\to2$ and
 NLO $2\to3$ (see the figures \ref{fig0}, \ref{fig9} and \ref{fig10p} for the LO and  NLO level
 sub-processes
 and the related discussion about  the very small contributions of  NLO level processes given in the equation (\ref{1p}) to cross-section
   in the last part of section (\ref{dis})).
The  Feynman diagrams are used to demonstrate the partonic
sector of the above processes. These are represented as the
following two LO partonic and one NLO sub-processes, respectively:
\begin{eqnarray}
q(\bar q)+g &\to&\gamma+ q(\bar q),\nonumber \\
q+\bar q &\to& \gamma+ g, \\
g+g &\to&\gamma+ q+\bar q, \nonumber \label{1}
\end{eqnarray}
namely "qg" (LO level), "qq" (LO level) and "gg" (NLO Level)
sub-processes. Also, the share of the other three NLO level sub-paratonic
contributions such as, quark-gluon and  quark-(anti)quark  
sub-processes:
\begin{eqnarray}
q(\bar q)+g &\to&\gamma+ q(\bar q)+g,\nonumber \\
 q(\bar q)+q(\bar q) &\to&\gamma+ q(\bar q)+  q(\bar q), \nonumber \\
q+\bar q &\to& \gamma+ g+g,  \label{1p}
\end{eqnarray}
are calculated separately (see figure 9, 10 and 11). It should be
noted that in the present work, we did not intend to calculate the
prompt single-photon production cross-section at the NLO level. Only
to show that there is not any double-counting between the $2 \to 2$
and $2 \to 3$ sub-processes, the $gg$ sub-process which is very
small compared to $qg$ and $qq$ sub-processes in the LO level, is
added to the all of prompt single-photon production cross-section
calculations. However, as we stated above the rest of NLO level
sub-processes are discussed in the section (\ref{dis}) and the
figure 11, which have much smaller effects to the differential cross
sections. It is also worth mentioning that for extracting the set of
$2 \to 3$ NLO Feynman diagrams according to the reference
\cite{olang,113,113-50,113-51,113-52,113-53}, one can use  the set
of $2\to 2$ LO Feynman diagrams by adding an additional parton
emitted from the initial or final partons (collinear factorization
approach).

In the case of NLO contributions, the real and virtual corrections should be considered to reach to the finite results and avoid the possible  UV-divergence. One can follow this issue for example in the recent review  by Konig \cite{Konig} as well as Passarino-Veltman reduction \cite{Veltman,Boehm} (also see, QCDLoop \cite{QCD} or LoopTools \cite{Loop} numerical libraries). 
Some divergences also appear because of the small $k_t$ ($<<\mu$) of the outgoing parton \cite{Boehm}. But since this parton is in the direction of outgoing photons, it is eliminated by excluding the above mentioned regions in our calculation
and also implementing isolated and separated cones in this computation. In this work, we
applied the same method as the reference \cite{jetphox,amin}   for the phase space cut.

To carry out the necessary calculations, one should  fix the
kinematics of these processes. Therefore,  the 4-momentum vectors of
the colliding protons are written as:
$$
P_1 = {\sqrt{s} \over 2} (1,0,0,1), \;\;\; P_2 = {\sqrt{s} \over 2}
(1,0,0,-1),
$$
where $s$ is the center-of-mass energy.

Using the modified Sudakov decomposition in the high energy
\cite{light-cone,Sudakov},
 \begin{equation}
\textbf{k}_i = x_i \textbf{P}_i + \textbf{k}_{i,t}, \;\;\;
    \label{eqs2}
    \end{equation}
then, the 4-momenta of the $i^{th}$ parton can be written as a
function of the transverse momenta, $k_{i,t}$, and the longitudinal
fraction of the momentum, $x_i$. These are assumed as inherited
parameters for a given parton. Considering the  sub-processes in the
equations (\ref{1}) and (\ref{1p}) and the conservation law of energy-momentum, the
relation between the Bjorken variable ($x$), the transfer momentum
($k_t^{i}$), and the rapidity ($y^{i}$) is obtained for the  $2\to2$ sub-processes as:
\begin{equation}
x_1 ={1\over \sqrt{s}} ( k_t^\gamma e^{y_{\gamma}}+ m_{t,1}
e^{y_{1}}),
\end{equation}
\begin{equation}
x_2 ={1\over \sqrt{s}} ( k_t^\gamma e^{-y_\gamma}+ m_{t,1}
e^{-y_{1}}) \label{eqs3}
\end{equation}
and for the  $2\to3$ sub-processes as:
\begin{equation}
x_1 ={1\over \sqrt{s}}( k_t^\gamma e^{y_{\gamma}}+ m_{t,1} e^{y_{1}}
+ m_{t,2} e^{y_{2}}),
\end{equation}
\begin{equation}
x_2 ={1\over \sqrt{s}}( k_t^\gamma e^{-y_{\gamma}}+ m_{t,1}
e^{-y_{1}} + m_{t,2} e^{-y_{2}}) \label{eqs4},
\end{equation}
where ${y_{\gamma}}$  and ${y_{1}}$ and ${y_{2}}$ are the photon
  and the outgoing partons rapidities, respectively. Also,
$m_{t,1}$ and $ m_{t,2}$ are the transverse masses  of outgoing
partons:
\begin{equation}
 m^2_{t,i}= m^2_i+ k^2_{t,i}.
\label{eqs5}
\end{equation}
Note that $m_{t,1}$  becomes $ k_{t,1}$ for the gluons
\cite{light-cone,Sudakov}.

So the incoming partons are off-shell, i.e., they carry transverse
momentum and should be described by the corresponding UPDF of the
$k_t$-factorization. As a result, the prompt single-photon
production differential cross-sections in the $k_t$-factorization frameworks
for the $2\longrightarrow 2$ sub-processes are \cite{amin,lip2007,lips0}:
$$
\sigma_{qg} (P_1+P_2\to q(\bar{q})+g\to \gamma+q(\bar{q})) = \int
{dk_{1,t}^2 \over k_{1,t}^2} \; {dk_{2,t}^2 \over k_{2,t}^2}
    dp_{\gamma,t}^2 \;  dy_{\gamma} \; dy_{1} \; {d\varphi_{1} \over 2\pi} \; {d\varphi_{2}\over 2\pi} \;{d\varphi_{\gamma}\over 2\pi} \;
    \; \times
    $$
      \begin{equation}
{|\mathcal{M}(q^*(\bar{q}^*)+g^* \rightarrow \gamma+q(\bar{q}))|^2
\over 16 \pi (x_1 x_2 s)^2} \; f_{q(\bar{q})}(x_1,k_{1,t}^2,\mu^2)
\; f_{g}(x_2,k_{2,t}^2,\mu^2),
    \label{eq251}
    \end{equation}
$$
\sigma_{q\bar q} (P_1+P_2\to q+\bar{q}\to \gamma+g) =  \int
{dk_{1,t}^2 \over k_{1,t}^2} \; {dk_{2,t}^2 \over k_{2,t}^2}
    dp_{\gamma,t}^2 \;  dy_{\gamma} \; dy_{1} \; {d\varphi_{1} \over 2\pi} \; {d\varphi_{2}\over 2\pi} \;{d\varphi_{\gamma}\over 2\pi} \;
    \; \times
    $$
      \begin{equation}
{|\mathcal{M}(q^*+\bar q^* \rightarrow \gamma+g)|^2 \over 16 \pi
(x_1 x_2 s)^2} \; f_{q}(x_1,k_{1,t}^2,\mu^2) \; f_{\bar
q}(x_2,k_{2,t}^2,\mu^2),
    \label{eq252}
    \end{equation}
and for the  $2\longrightarrow3$ sub-process is,
 $$
\sigma_{gg} (P_1+P_2\to g+g\to \gamma+q+\bar{q}) =  \int {dk_{1,t}^2
\over k_{1,t}^2} \; {dk_{2,t}^2 \over k_{2,t}^2} \;
    dp_{\gamma,t}^2 \; dp_{1,t}^2 \; dy_{\gamma} \; dy_{1} \; dy_2
    \; \times
    $$
    $$
{d\varphi_{1} \over 2\pi} \; {d\varphi_{2} \over 2\pi} \;
{d\varphi_{\gamma} \over 2\pi} \; {d\psi_{1} \over 2\pi}
    \times
    $$
    \begin{equation}
{|\mathcal{M}(g^*+g^*\to \gamma+q+\bar{q})|^2 \over 256 \pi^3 (x_1
x_2 s)^2} \; f_{g}(x_1,k_{1,t}^2,\mu^2) \;
f_{g}(x_2,k_{2,t}^2,\mu^2),
    \label{eq261}
    \end{equation}
where the total differential cross-sections can be written as:
$$
\sigma_{T}=\sigma_{qg}+\sigma_{q\bar q}+\sigma_{gg}
$$
The various dispute about the consideration of above
$2\longrightarrow 2$ and $2\longrightarrow 3$ sub-processes, as it
was mentioned in the introduction will be discussed in the section
III \cite{amin,KIMBER,lips0,lip2007,W/Z-LHCb,lips5}.

For calculating the single and double-differential cross-section of
prompt single-photon production in the equations (\ref{eq251}),
(\ref{eq252}) and (\ref{eq261}), the matrix element squared,
$|\mathcal{M}|^2 $,  of the sub-processes of the equation (1)  must
be calculated. By considering that the incoming partons must be
off-shell, the matrix element, $\mathcal{M} $, for the sub
processes, which introduced in these equations are defined in the
Appendix A and, their Feynman diagrams are displayed in the figure
\ref{fig0}. Some of the squared matrix elements are also given in
the references \cite{lip2007,PRD75,lips0}. The  VEGAS algorithm is
considered for performing  the multidimensional integration
   of the  total differential cross-section in the  equations (\ref{eq251}) to (\ref{eq261}).
\subsection{The choice of UPDF}
\label{UPDF} The UPDF can be directly generated from the PDF, by
using different prescriptions. In this work, we  use three
approaches, the so called KMR \cite{KMR}, LO  and  NLO MRW
\cite{MRW}, to obtain the UPDF from the corresponding PDF  and substitute them  in the equations (\ref{eq251}), (\ref{eq252}) and
(\ref{eq261}).

The KMR UPDF are  generated such that the partons evolve from the
starting PDF parametrization up to the scale $k_t$ according to the
DGLAP evolution equations \cite{KMR}. In the KMR method, the partons
emit in the single evolution ladder (carrying only the $k_t^2$
dependency) and get convoluted with the second scale, $\mu^2$, at
the hard process. The $k_t$  is assumed to be depend on  the scale
$\mu^2$, without any real emission, and by summing over the virtual
contributions via the Sudakov form factor, $T_a(k_t^2,\mu^2)$. So,
the general form of the KMR UPDF are:
   \begin{equation}
f_a(x,k_t^2,\mu^2) = T_a(k_t^2,\mu^2)\sum_{b=q,g} \left[
{\alpha_S(k_t^2) \over 2\pi} \int^{1-\Delta}_{x} dz P_{ab}^{(LO)}(z)
b\left( {x \over z}, k_t^2 \right) \right] , \label{eq56}
    \end{equation}
where $T_a(k_t^2,\mu^2)$ are :
   \begin{equation}
T_a(k_t^2,\mu^2) = exp \left( - \int_{k_t^2}^{\mu^2} {\alpha_S(k^2)
\over 2\pi} {dk^{2} \over k^2} \sum_{b=q,g} \int^{1-\Delta}_{0} dz'
P_{ab}^{(LO)}(z') \right). \label{eq5}
    \end{equation}
$T_a$ are considered to be unity for $k_t>\mu$.  $\Delta$  is
proposed  for the soft gluon and the quark radiations.

To determine $\Delta$, the angular ordering constraint is imposed.
The angular ordering originates from the color coherence effects of
the gluon radiations \cite{KMR}. So $\Delta$ is:
$$ \Delta = {k_t \over \mu + k_t} .$$
There is also the strong ordering constraint, i.e., $ \Delta = {k_t
\over \mu  } $  \cite{WattWZ,amin,O,kimber2}.
$P_{ab}^{(LO)}(z)$ are the LO splitting functions
\cite{co}.

The LO MRW UPDF are similar to the KMR ones, but with the different
treatment of the  angular ordering constraint. In this approach
angular ordering constraint correctly imposed only on the  soft
gluon radiations, i.e., the diagonal splitting functions
$P_{qq}(z)$ and $P_{gg}(z)$ \cite{MRW}. So, the LO MRW prescription
is written as:
    $$
f_q^{LO}(x,k_t^2,\mu^2)= T_q(k_t^2,\mu^2) {\alpha_S(k_t^2) \over
2\pi} \int_x^1 dz \left[ P_{qq}^{(LO)}(z) {x \over z} q \left( {x
\over z} , k_t^2 \right) \Theta \left( {\mu \over \mu + k_t}-z
\right) \right.
    $$
    \begin{equation}
\left. + P_{qg}^{(LO)}(z) {x \over z} g \left( {x \over z} , k_t^2
\right) \right], \label{eq7}
    \end{equation}
with
    \begin{equation}
T_q(k_t^2,\mu^2) = exp \left( - \int_{k_t^2}^{\mu^2} {\alpha_S(k^2)
\over 2\pi} {dk^{2} \over k^2} \int^{z_{max}}_{0} dz'
P_{qq}^{(LO)}(z') \right), \label{eq8}    \end{equation} for the
quarks and
    $$
f_g^{LO}(x,k_t^2,\mu^2)= T_g(k_t^2,\mu^2) {\alpha_S(k_t^2) \over
2\pi} \int_x^1 dz \left[ P_{gq}^{(LO)}(z) \sum_q {x \over z} q
\left( {x \over z} , k_t^2 \right)
    \right.$$
    \begin{equation}
\left. + P_{gg}^{(LO)}(z) {x \over z} g \left( {x \over z} , k_t^2
\right) \Theta \left( {\mu \over \mu + k_t}-z \right)
    \right], \label{eq9}
    \end{equation}
with
    \begin{equation}
T_g(k_t^2,\mu^2) = exp \left( - \int_{k_t^2}^{\mu^2} {\alpha_S(k^2)
\over 2\pi} {dk^{2} \over k^2}
    \left[ \int^{z_{max}}_{z_{min}} dz' z' P_{gg}^{(LO)}(z')
+ n_f \int^1_0 dz' P_{qg}^{(LO)}(z') \right] \right) , \label{eq10}
    \end{equation}
for the gluons. In the equations (\ref{eq8}) and (\ref{eq10}),
$z_{max}=1-z_{min}=\mu/(\mu+k_t)$ \cite{WATT}.

The Martin et al.   \cite{MRW} also  proposed   the NLO MRW
formalism. This method is based on the  DGLAP  evolution equation,
utilizing the NLO PDF and  the corresponding splitting functions
\cite{MRW}. The general form of the NLO MRW UPDF is:
    $$
f_a^{NLO}(x,k_t^2,\mu^2)= \int_x^1 dz T_a \left( k^2={k_t^2 \over
(1-z)}, \mu^2 \right) {\alpha_S(k^2) \over 2\pi}
    \sum_{b=q,g} \tilde{P}_{ab}^{(LO+NLO)}(z)
    $$
    \begin{equation}
\times b^{NLO} \left( {x \over z} , k^2 \right) \Theta \left(
1-z-{k_t^2 \over \mu^2} \right),
    \label{eq11}
    \end{equation}
with the "extended" $NLO$ splitting functions,
$\tilde{P}_{ab}^{(i)}(z)$,  being defined as:
    \begin{equation}
\tilde{P}_{ab}^{(LO+NLO)}(z) = \tilde{P}_{ab}^{(LO)}(z) + {\alpha_S
\over 2\pi}
    \tilde{P}_{ab}^{(NLO)}(z),
    \label{eq12}
    \end{equation}
and
    \begin{equation}
\tilde{P}_{ab}^{(i)}(z) = P_{ab}^{i}(z) - \Theta (z-(1-\Delta))
\delta_{ab} F^{i}_{ab} P_{ab}(z),
    \label{eq13}
    \end{equation}
where $i=0$ and $1$ stand for the LO and the NLO, respectively.
Also, the angular ordering constraint is defined via the $\Theta
(z-(1-\Delta))$ in  which $\Delta$ is defined as \cite{MRW}:
    $$ \Delta = {k\sqrt{1-z} \over k\sqrt{1-z} + \mu}.$$
Finally, the Sudakov form factors in the NLO MRW approach are defined as:
    \begin{equation}
T_q(k^2,\mu^2) = exp \left( - \int_{k^2}^{\mu^2} {\alpha_S(q^2)
\over 2\pi} {dq^{2} \over q^2} \int^1_0 dz' z' \left[
\tilde{P}_{qq}^{(0+1)}(z') + \tilde{P}_{gq}^{(0+1)}(z') \right]
\right) ,
    \label{eq14}
    \end{equation}
    \begin{equation}
T_g(k^2,\mu^2) = exp \left( - \int_{k^2}^{\mu^2} {\alpha_S(q^2)
\over 2\pi} {dq^{2} \over q^2} \int^1_0 dz' z' \left[
\tilde{P}_{gg}^{(0+1)}(z') + 2n_f\tilde{P}_{qg}^{(0+1)}(z') \right]
\right) .
    \label{eq15}
    \end{equation}

Note that the integral intervals for $dk_{t}$ integration, i.e., the
equations (\ref{eq251}), (\ref{eq252}), are $(0,\infty)$, so one can
choose an upper limit for these integrations, say $k_{i,max}$,
several times bigger than the hard scale $\mu$. Also,
$k_{t,min}=\mu_0\sim1$ $ GeV$,  is considered as the lower limit
that separates the non-perturbative and the perturbative regions, by
assuming that,
\begin{equation}
{1\over {k_t^2}}\; f_a(x,k_t^2,\mu^2) {\huge |}_{k_t<\mu_0}  =
{1\over \mu_0^2}\;a(x,\mu_0^2)\;T_a(\mu_0^2,\mu^2).
    \label{eq0025}
\end{equation}
Eventually, the density of patrons are constant for $k_t<\mu_0$ at
fix $x$ and $\mu$ \cite{MRW}.
 For the above calculations, we  use the LO-MMHT2014 PDF libraries \cite{MMHT}
  for the  KMR  and the  LO MRW  UPDF schemes, and the NLO-MMHT2014 PDF libraries \cite{MMHT} for
 the  NLO MRW formalism.
   A more complicated extrapolation of the contribution from
$k_t\leq \mu_0$, which ensures the continuity of
$f_a(x,k_t^2,\mu^2)\mid_{k_t\leq \mu_0}$, is given in the references
\cite{MRW,WattWZ}. This is done by a polynomial expansion of
$f_a(x,k_t^2,\mu^2)$ for the requirement of $k_t^2$ behavior of the
UPDF at the small $k_t^2$.
\subsection{The experimental conditions}
\label{cond} The fragmentation contributions \cite{au} are
neglected in our calculations, since  the isolation cut application
\cite{D02005,CDF2017} reduces these contributions to less than 10
percent of the   cross-section. Furthermore,  the isolation
cuts and additional conditions which preserve our calculations from
divergences were specially discussed in the references \cite{amin,
lips8} as follows: The isolated-cone is responsible for
distinguishing the "non-prompt decay photons" from the
prompt-photons. This constraint requires that the transverse energy
$E_t^{Hadron}$ (in a cone with the angular radius $ R\geq
\sqrt{(\eta-\eta^\gamma)^2\;+\;(\phi-\phi_\gamma)^2}$) to be less
than a few GeV according to each experiment, where $\eta$ and $\phi$
are the pseudorapidities and the azimuthal angle plane of the hadron
and $R\sim 0.4-0.7$, depending on each experimental collaborations.
 Other constraints, such as the $p_t$-threshold of prompt-photons, the pseudo-rapidity
 regions, etc, are imposed  according to the settings of the individual
 experiments.
\section{Results and discussions}
\label{sec:results}
\subsection{The numerical results}
\label{num}
 We  perform a set of numerical calculations   for the production of the single-photon at the LHC and
TEVATRON colliders, using the equations  (\ref{eq251}),
(\ref{eq252}) and (\ref{eq261}), within the $k_t$-factorization and
the different UPDF  approaches. The results are separated into the 1.96,
7 and 13 TeV center-of-mass energy ($E_{CM}=\sqrt{s}$), in
accordance with the specifications of  existing experimental data
\cite{D02005,CDF2017,Atlas2017,cms}. The ratio of our prompt
single-photon production differential cross-sections calculation to
those of experimental data at each bin ($\cal R$) are separately
presented (see the panels (d) of our figures). We also compare our
results with those of Monte Carlo simulations which were introduced
in the introduction.

In the figures \ref{fig1} and  \ref{fig2} the reader is presented
with the double differential cross-sections for the production of a
single-photon ($d^2\sigma / dp_{\gamma,t}dy$) as a function of its
transverse momentum ($p_{\gamma,t}$) at $E_{CM} = 1.96$ TeV . These
results are in accordance with the experimental data of the D0
\cite{D02005} and CDF \cite{CDF2017} collaborations by applying
specific experimental conditions. Similarly, in  the figures
\ref{fig3} to \ref{fig6} and \ref{fig7}, the same  results are
present for ATLAS measurements \cite{Atlas2017} at $E_{CM} =
13\;TeV$ and CMS collaboration  at $E_{CM} = 7$ TeV, respectively.

Within each of these figures, the panel (a) illustrates our results
from the utilization of the KMR UPDF, while the panels (b) and (c)
exhibit those of the LO and NLO MRW UPDF, respectively. In each of
these panels, the contributions from the partonic sub-processes are
demonstrated as follows: the red dashed histograms for   $q(\bar
q)+g \to\gamma+ q(\bar q)$, the green dotted lines for   $q+\bar q
\to \gamma+g$ and  the blue dotted-dash lines for   $g+g \to
\gamma+q+\bar q$. The sum of these contributions for
each approaches, i.e., KMR  and  LO and NLO MRW, are shown with the
black solid lines. The corresponding uncertainty regions (see the
blue hatched areas in the panel (a)) are calculated, only for the KMR framework, by manipulating
the hard-scale $\mu$ by a factor of $2$, i.e., ${1\over 2}\mu$ to
$2\mu$. A comparison is also made
between all three UPDF approaches  and the results of others
theoretical calculations, as explained in the corresponding captions
of   figures (see the panels (d)). Since the experimental results
are given at different bins, only in the panels (d) in which the our
final  differential cross-sections are presented, we average over
our results corresponding to each experimental bin.

In the ATLAS collaboration data \cite{Atlas2017}, the corresponding
measurements were carried out within the following rapidity regions:
\begin{enumerate}
    \item $|y|<0.6$,
    \item $0.6<|y|<1.37$,
    \item $1.56<|y|<1.81$,
    \item $1.81<|y|<2.37$,
\end{enumerate}
which presented in the figures \ref{fig3} to \ref{fig6},
respectively.
\subsection{Discussions}
\label{dis} In the panel (a) of  the figures \ref{fig1} and
\ref{fig2}, it is     clear that the $qg$ sub-process  has large
contribution, which  confirms  the behavior that was  reported in
the references \cite{Atlas2017,D02005}. Also, in the small
$p_{\gamma,t}$, the contribution of gg sub-processes is larger than
those of  qq. However, their  values   become close together around
$p_{\gamma,t} = 30-50$ GeV, but in the large photon  momentum
transfer, the portion of qq contribution becomes enhanced with
respect to the gg sub-process. The same behavior is observed in the
panels (b) and (c), which   corresponds to   utilizing the LO and
NLO MRW $k_t$-factorization frameworks. Although,  the qg
sub-processes still have a  dominant  contribution, but the crossing
interval of gg and qq curves   behave differently in each framework.
In the panel (d), the final result of three approaches are displayed
for the comparison related to the experimental data \cite{D02005}.
It becomes clear that, there is no significant difference between
each $k_t$-factorization scheme. Despite differences in the behavior
of the partonic sub-processes, these results are relatively similar
and in agreement with the experimental findings. However, the small
differences which originated from the application of above three
schemes, are the effects of different angular ordering
constraints. The ratios of our results to those of experimental data
are demonstrated at the bottom of these panels ($\cal R$). It is
observed that at the small photon transverse momentum regions there
are good agreement between the  present calculation and the data.

Similar calculations are also   made for $E_{CM} = 13$
 TeV, corresponding to the experimental data of the reference
\cite{Atlas2017}. The results are presented in the figures
\ref{fig3} to \ref{fig6}. Here, the rapidity region is separated
into 4 sectors. The general behaviors of the contributing
sub-processes are similar to the $1.96$ TeV case. With increasing
the center-of-mass energy of the hadronic collision, the results
coming from the NLO MRW framework maintain their relative success in
 predicting of the experimental data. On the other hand, a
tangible difference in the precision of the LO MRW and KMR
predictions can be observed.

By moving up between the rapidity regions, it is evident that by
increasing the rapidity, the crossing intervals  between the qq and
gg sub-processes are moved to the small transverse momentum regions.
However, they completely disappear in the $1.81<|y|<2.37$ rapidity
interval. Actually, the portion  of qq contribution is enhanced with
respect to the gg  sub-processes, while the qg contribution is
strongly dominated.

Similar conclusion as above can be made about  $\cal R$ in the
panels (d) of these figures with this difference that the NLO MRW
differential cross-sections are closer to the data.

The same discussion, as the one given above, can be made about the
figure \ref{fig7}, but with this difference that the contribution of
gg becomes larger in the small photon transverse  momentum. This is
expected since the center of mass energy is much high with respect
to the figures \ref{fig1} and  \ref{fig2}. Note that in the figure
\ref{fig3} to \ref{fig6},   the ATLAS data probing the photon
transverse  momentum form much lager values, i.e., $125$ GeV. On the
other hand, as it was pointed out in the reference \cite{amin}, in
the fragmentation region ($p_{\gamma,t}\leq 30$ GeV ) the LO MRW can
support the data much   better.

In all of our figures, we   also present  the results of theoretical
prompt single-photon production differential cross-section
calculations of different theoretical groups mentioned in the
introduction, such as, JETPHOX, SHERPA, PYTHIA and  MCFM. A good
agreement can be seen between our different $k_t$-factorizations
calculations and these collinear factorization simulation methods.
Similar to our photon pair  report they are in the $k_t$-angular
ordering constraint bound \cite{amin}

It should be pointed out that beside the NLO gg-fusion process  (see
the equation (\ref{1})), the channels $q+g \to q +g+ \gamma$
(similar to its Compton LO counterpart), $q^*+\bar q^* \to \gamma+q+
\bar q $ and $q^*+\bar q^* \to \gamma+g+ g $ also appears at the NLO
level. On the other hand,  when considering the inclusive photon
production, the NLO diagrams can contain collinearly enhanced
productions that should be factorized properly
\cite{olang,113,113-50,113-51,113-52,113-53,light-cone}. Therefore,
to increase the accuracy of the  cross-section calculations, the
above sub-processes are also computed at NLO  level separately and
their  diagrams are  shown in the figures \ref{fig9} and
\ref{fig10p} (the mathematical formula of  the cross-sections for these
processes are very similar to that of equation (\ref{eq261})). The
result of this calculation is presented in the figure \ref{fig10}
which is
 based on the specific condition that depicted at each panel. However, the shares
 of these
 NLO    sub-processes are clearly much less than   LO level and
 gluons-fusion NLO
   sub-processes, but in over all  these portions  make the results better in the small photon transverse momentum. In the figure \ref{fig12}, the different contributions of the LO and NLO levels to the differential cross-sections as in the figure \ref{fig10} are shown separately. It is observed that  the NLO sub-processes have very small contributions and the LO level  mainly   contribute to the cross-sections.  However for very small photon transverse momentum the gluons-fusion has sizeable contribution to the cross-sections, as it is pointed out by the CMS collaboration \cite{cms}. We did not   calculate the NLO virtual corrections to the LO quark-gluon process. But they could be the same order as the quark-gluon NLO process, which    we have already calculated. On the other hand, one should note that we are not working in the collinear factorization frame work, but the $k_t$-factorization. There are quark, antiquark, and gluon degrees of freedom inside of the UPDF such that at LO level one can argue that, e.g. the KMR is a semi-NLO approach.
   
As it was pointed out in the introduction, we should also discuss
about the possible double-counting between our $2 \to 2$ and  $2\to
3$ sub-processes, which were presented in our results.  In some of
the references \cite{lips,lips0,lips1,lips5,lips7,lips8,lip2007},
the $2\to 3$ sub-process is neglected, or if it  is considered, only
the sea-quarks contributions in the UPDF are omitted on the basis of
double-counting, e.g., in the region where the transverse momentum
of one of the parton is as large as the hard scale and the
additional parton is highly separated in the rapidity, from the hard
process (multi-Regge region), while the additional emission in the
$2\to 3$  sub-process  was subtracted. But, in general, one should
consider the KMR or MRW UPDF in the  $k_t$-factorization
calculations corresponds to the non-normalized probability
functions. They are used as the weight of the given transition
amplitudes (the matrix elements in these cases). The transverse
momentum dependence of the UPDF come from considering all possible
splittings up to and including the last splitting, see the
references \cite{KMR,WattWZ,WATT,KMR1}, while the evolution up to
the hard scale without a change in the $k_t$, due to virtual
contributions, is encapsulated in the Sudakov-like survival form
factor. Therefore, all splittings and real emissions of the partons,
including the last emission, are factorized in the function $f_g(x,
k_t^2, \mu^2)$, as its definition. The last emission from the
generated  UPDF, can not be disassociated and to be count as the
part of the $2\to 3$ diagrams \cite{W/Z-LHCb,KIMBER}. Recently, a
detailed investigation of the above possible double counting is also
reported in the reference \cite{Guiot}, which confirms our
conclusions. One should also note that the UPDF have to satisfy the
condition given in the identity equation:
\begin{equation}
    {xa(x,Q^2) \simeq \int^{Q^2} {{dk_t^2} {k_t^{-2}}}} f(x,k^2_t,Q^2).
    \label{eq0065}
\end{equation}
So any changes in the UPDF certainly affect the original PDF
definitions and it alters the whole formalism.

However, as it was discussed in  the argument in connection to our
results, depending on the value of the photon transverse momentum
the $2\to 2$ (large $p_{\gamma,t}$) or $2\to 3$ process, i.e., only gluons-fusion (small
$p_{\gamma,t}$) make the main contributions to the prompt
single-photon production differential cross-sections. To check our
results, we also made an approximation and modified our UPDF
according to the reference \cite{lips5}  and found that the results
are still in the uncertainty bound region.

In the small x limit, see the reference \cite{amin} and the
reference therein,  the off-shell matrix element satisfies the gauge
invariance. However,  there are   reggeization methods to evaluate
the off-shell quark density matrix elements which inherently satisfy
the gauge invariance in all regions \cite{lips1} or       the vertex
modification  \cite{KUTAK1}, using the auxiliary photons and quarks.
To be sure about the above problem,   we checked our result
numerically similar to reference \cite{lips0} as well as imposing
the on-shell matrix elements but with the $k_t$-factorization
dynamics \cite{KMR1,kimber2}. We did not find much difference
between the off-shell and on-shell matrix elements calculations of
differential cross-sections, especially in the large photon momentum
transverse, i.e., $\geq 30$ GeV.
\section{Conclusion}
\label{sec:conc} Throughout this work,  using the  UPDF of
$k_t$-factorization, i.e., the KMR and the LO and NLO MRW
frameworks, we calculated the   rate of  prompt single-photon
production at the LHC and TEVATRON colliders for the center-of-mass
energies of 7 and 13 and 1.96 TeV, respectively. We  compared our
numerical results against each other and those of the experimental
data from the D0, CDF, CMS and ATLAS collaborations. Our aim was
generally to illustrate capability of the above UPDF to  describe
the experimental measurements but do not to increase the precision
of such predictions, especially compared to well developed collinear
frameworks that are currently   used by different collaborations. It
was demonstrated that the despite of the simplicity  of our model,
(on the numerical sense), the UPDF of $k_t$-factorization are able
to successfully describe the experimental measurements. The further
increase in the precision of our calculations is achievable by
adding the higher-order and radiation corrections into the
matrix elements as well as providing more accurate UPDF via
undergoing a complete phenomenological global fit to the existing
deep inelastic data \cite{WattWZ,O}. Also, for increasing the
accuracy, we added the other NLO level sub-processes contributions
to the differential cross-sections by using the Feynman rules. These
process increased the precision of result only in the small
$p_{t,\gamma}$.

As it was mentioned in the sections I and II,  the difference in
calculating the differential cross-sections with the on-shell or the
off-shell matrix element, can affect the results at the small photon
momentum transverse regions ($\leq 30$ GeV) and one can conclude
that in this region the gauge invariance of the off-shell matrix
elements become important. So, on this basis, we are interested to
repeat present calculations, using the method introduced in the
reference \cite{KUTAK1}, concerning the gauge invariance of the off
shell matrix element in whole x region. It would be also interesting
to investigate the influence of different modifications of UPDF and
to use the various angular ordering constraints to examine their
behavior  in order to check the possible double-counting as well as
to consider the differential method to generate the UPDF
\cite{O,MRW,WATT,WattWZ,GS}. We hope to report these points in our
future works.
\begin{acknowledgments}
MM would like to acknowledge the Research Council of University of
Tehran   for the grants provided for him. RAN sincerely thanks M.
Kimber and A. Lipatov for their valuable discussions and comments. We would like to acknowledge the research
support of the Iran National Science
Foundation (INSF) for their  grants.
\end{acknowledgments}
\appendix
\section{}
The various matrix elements of the processes defined in the
equations
(1) and (2) and the figures \ref{fig0},  \ref{fig9} and \ref{fig10p}, respectively, are given as:\\
1)\;$q^*(\bar q^*)+g^* \to \gamma+q (\bar q)$ (LO, figure
\ref{fig0}):
$$
\mathcal{M}_{a}=  e  g \bar{U}(p_1) t^a \epsilon_\mu(k_2)
\epsilon_\nu(p)
 (\gamma^{\mu}  {\slashed{k_1}-\slashed{p}+m \over
  (k_1-p)^2-m^2} \gamma^{\nu} + \gamma^{\nu}  {\slashed{k_1}+\slashed{k_2}+m \over
  (k_1+k_2)^2-m^2} \gamma^{\mu} )  U(k_1),
$$
2)\; $q^*+\bar q^* \to \gamma+g$ (LO, figure \ref{fig0}):
$$
\mathcal{M}_{b}=  e  g \bar{U}(k_2) t^a \epsilon_\mu(p_1)
\epsilon_\nu(p)  (\gamma^{\mu}
 {\slashed{k_1}-\slashed{p}+m \over
  (k_1-p)^2-m^2} \gamma^{\nu} + \gamma^{\nu}  {\slashed{k_1}-\slashed{p_1}+m \over
  (k_1-p_1)^2-m^2} \gamma^{\mu} )  U(k_1)
$$
3)\; $g^*+g^* \to \gamma+q+\bar q$ \cite{lips0} (NLO, figure
\ref{fig0}):
$$
\mathcal{M}_1 =  e  g^2    \bar{U}(p_1)  t^{\alpha} \gamma^{\mu}
 \epsilon_\mu(k_1)    {\slashed{p_1}-\slashed{k_1}+m \over
  (p_1-k_1)^2-m^2}   \gamma^{\xi}   \epsilon_\xi(p) {\slashed{k_2}-\slashed{p_2}+m \over
(k_2-p_2)^2 - m^2}   t^{\beta}   \gamma^{\nu} \epsilon_\nu(k_2)
U(p_2),
    $$
 $$
\mathcal{M}_2 =  e  g^2    \bar{U}(p_1)  t^{\beta} \gamma^{\nu}
 \epsilon_\nu(k_2)   {\slashed{p_1}-\slashed{k_2}+m \over
  (p_1-k_2)^2-m^2}   \gamma^{\xi}   \epsilon_\xi(p) {\slashed{k_1}-\slashed{p_2}+m \over
(k_1-p_2)^2 - m^2}   t^{\alpha}   \gamma^{\mu} \epsilon_\mu(k_1)
   U(p_2),
    $$
$$
\mathcal{M}_3 =  e  g^2   \bar{U}(p_1)   t^{\alpha} \gamma^{\mu}
 \epsilon_\mu(k_1)    {\slashed{p_1}-\slashed{k_1}+m \over
  (p_1-k_1)^2-m^2}   t^{\beta}  \gamma^{\nu}   \epsilon_\nu(k_2) {-\slashed{p_2}-\slashed{p}+m \over
(-p-p_2)^2 - m^2}    \gamma^{\xi} \epsilon_\xi(p)    U(p_2),
    $$
$$
\mathcal{M}_4 =  e  g^2   \bar{U}(p_1)   t^{\beta} \gamma^{\nu}
 \epsilon_\nu(k_2)    {\slashed{p_1}-\slashed{k_2}+m \over
  (p_1-k_2)^2-m^2}   t^{\alpha}  \gamma^{\mu}   \epsilon_\mu(k_1) ;\ {-\slashed{p_2}-\slashed{p}+m \over
(-p-p_2)^2 - m^2}    \gamma^{\xi} \epsilon_\xi(p)    U(p_2),
    $$
$$
             \mathcal{M}_5 =  e  g^2   \bar{U}(p_1) \gamma^{\xi}  \epsilon_\xi(p)    {\slashed{p_1}+\slashed{p}+m \over
  (p_1+p)^2-m^2}   t^{\alpha}  \gamma^{\mu}   \epsilon_\mu(k_1) {\slashed{k_2}-\slashed{p_2}+m \over
(k_2-p_2)^2 - m^2}    t^{\beta}   \gamma^{\nu} \epsilon_\nu(k_2)
   U(p_2),
    $$
   $$
         \mathcal{M}_6 =  e  g^2   \bar{U}(p_1) \gamma^{\xi}  \epsilon_\xi(p)    {\slashed{p_1}+\slashed{p}+m \over
  (p_1+p)^2-m^2}   t^{\beta}  \gamma^{\nu}   \epsilon_\nu(k_2) {\slashed{k_1}-\slashed{p_2}+m \over
(k_1-p_2)^2 - m^2}    t^{\alpha}   \gamma^{\mu} \epsilon_\mu(k_1)
   U(p_2),
    $$
$$
             \mathcal{M}_7 = - e  g^2   \bar{U}(p_1) \gamma^{\rho}  C^{\mu \nu \rho}(k_1,k_2,-k_1-k_2)  {\epsilon_\mu(k_1)   \epsilon_\nu(k_2) \over (k_1+k_2)^2}   f^{abc} t^c {-\slashed{p_2}-\slashed{p}+m \over
  (-p_2-p)^2-m^2}   \gamma^{\xi} \epsilon_\xi(p)
   U(p_2),
    $$
$$
\mathcal{M}_8 = - e  g^2   \bar{U}(p_1) \gamma^{\xi} \epsilon_\xi(p)
 {\slashed{p_1}+\slashed{p}+m \over
  (p_1+p)^2-m^2} \gamma^{\rho}  C^{\mu \nu \rho}(k_1,k_2,-k_1-k_2)  {\epsilon_\mu(k_1)   \epsilon_\nu(k_2) \over (k_1+k_2)^2}   f^{abc} t^c
   U(p_2).
$$
4)\;   $q^*(\bar q^*)+g^* \to \gamma+q (\bar q)+g $, (NLO, figure
\ref{fig9}):
$$
\mathcal{M}_1 =  e  g^2    \bar{U}(p_1)   \gamma^{\xi}
\epsilon_\xi(p)  {\slashed{p_1}+\slashed{p}+m \over(p_1+p)^2 - m^2}
t^{\alpha} \gamma^{\mu} \epsilon_\mu(p_2)
{\slashed{k_1}+\slashed{k_2}+m \over(k_1+k_2)^2 - m^2}t^{\beta}
\gamma^{\nu} \epsilon_\nu(k_2)U(k_1),
 $$
 $$
\mathcal{M}_2 =  e  g^2    \bar{U}(p_1) t^{\alpha} \gamma^{\mu}
\epsilon_\mu(p_2)   {\slashed{p_1}+\slashed{p_2}+m \over(p_1+p_2)^2
- m^2}  \gamma^{\xi}  \epsilon_\xi(p) {\slashed{k_1}+\slashed{k_2}+m
\over(k_1+k_2)^2 - m^2}t^{\beta}   \gamma^{\nu}
\epsilon_\nu(k_2)U(k_1),
    $$
$$
\mathcal{M}_3 =  e  g^2    \bar{U}(p_1) t^{\alpha} \gamma^{\mu}
\epsilon_\mu(p_2)   {\slashed{p_1}+\slashed{p_2}+m \over(p_1+p_2)^2
- m^2}t^{\beta}   \gamma^{\nu}
\epsilon_\nu(k_2){\slashed{k_1}-\slashed{p}+m \over(k_1-p)^2 -
m^2}\gamma^{\xi}  \epsilon_\xi(p) U(k_1), $$
$$
\mathcal{M}_4 =  e  g^2    \bar{U}(p_1)t^{\beta}   \gamma^{\nu}
\epsilon_\nu(k_2){\slashed{k_2}-\slashed{p_1}+m \over(k_2-p_1)^2 -
m^2}\gamma^{\xi}  \epsilon_\xi(p) {\slashed{k_1}-\slashed{p_2}+m
\over(k_1-p_2)^2 - m^2}t^{\alpha} \gamma^{\mu} \epsilon_\mu(p_2)
U(k_1), $$
$$
\mathcal{M}_5 =  e  g^2    \bar{U}(p_1)t^{\beta}   \gamma^{\nu}
\epsilon_\nu(k_2){\slashed{k_2}-\slashed{p_2}+m \over(k_2-p_2)^2 -
m^2}t^{\alpha} \gamma^{\mu} \epsilon_\mu(p_2)
{\slashed{k_1}-\slashed{p}+m \over(k_1-p)^2 - m^2}\gamma^{\xi}
\epsilon_\xi(p)   U(k_1),
    $$
   $$
         \mathcal{M}_6 = - e  g^2   \bar{U}(p_1) \gamma^{\xi} \epsilon_\xi(p){\slashed{p_1}+\slashed{p}+m \over(p_1+p)^2-m^2} \gamma^{\rho}  C^{\mu \nu \rho}(k_1,k_2,p)  {\epsilon_\mu(p_2)   \epsilon_\nu(k_2) \over (k_1-p_1)^2}   f^{abc} t^c U(k_1),
    $$
$$
\mathcal{M}_7 =  e  g^2    \bar{U}(p_1)\gamma^{\xi}
\epsilon_\xi(p){\slashed{k_1}+\slashed{p}+m \over(k_1+p)^2 - m^2}
t^{\beta}   \gamma^{\nu}
\epsilon_\nu(k_2){\slashed{k_1}-\slashed{p_2}+m \over(k_1-p_2)^2 -
m^2}t^{\alpha} \gamma^{\mu} \epsilon_\mu(p_2)    U(k_1),
    $$
$$
\mathcal{M}_8 = - e  g^2   U(k_1)  \gamma^{\xi}
\epsilon_\xi(p){\slashed{k_1}-\slashed{p}+m \over(k_1-p)^2-m^2}
\gamma^{\rho}  C^{\mu \nu \rho}(k_1,k_2,p)  {\epsilon_\mu(p_2)
\epsilon_\nu(k_2) \over (k_1-p_1)^2}   f^{abc} t^c\bar{U}(p_1),
$$
5)\; $q^*+\bar q^*  \to \gamma+q+ \bar q $ (NLO, figure
\ref{fig10p})
$$
\mathcal{M}_{a.1} =  e  g^2    \bar{U}(p_2)   \gamma^{\mu}
\epsilon_\mu(p)  {\slashed{p_2}+\slashed{p}+m \over(p_2+p)^2 - m^2}
U(p_1) t^{\alpha} \gamma^{\xi} {g^{\xi \nu} \delta^{\alpha \beta}
\over (k_1+k_2)^2} t^{\beta} \bar{U}(k_2) \gamma^{\nu} U(k_1) ,
 $$
  $$
\mathcal{M}_{a.2} =   e  g^2    \bar{U}(p_2)   \gamma^{\mu}
\epsilon_\mu(p)  {\slashed{p_2}+\slashed{p}+m \over(p_2+p)^2 - m^2}
\bar{U}(k_2) t^{\beta}  \gamma^{\nu} {g^{\xi \nu} \delta^{\alpha
\beta} \over (k1-p1)^2} U(p_1) t^{\alpha} \gamma^{\xi}   U(k_1) ,
    $$
$$
\mathcal{M}_{a.3} =  e  g^2    \bar{U}(p_2)  t^{\alpha} \gamma^{\xi}
U(p_1) {g^{\xi \nu} \delta^{\alpha \beta} \over (p_2+p_1)^2}
\bar{U}(k_2) t^{\beta}  \gamma^{\nu}  {\slashed{k_1}-\slashed{p}+m
\over(k_1-p)^2 - m^2}    \epsilon_\mu(p) \gamma^{\mu}   U(k_1) , $$
$$
\mathcal{M}_{a.4} =  e  g^2    U(p_1) \gamma^{\xi} \epsilon_\xi(p)
{\slashed{p_1}+\slashed{p}+m \over(p_1+p)^2 - m^2}
\bar{U}(p_2)\gamma^{\mu}  t^{\alpha} {g^{\mu \nu} \delta^{\alpha
\beta} \over (k_2+k_1)^2} t^{\beta} \bar{U}(k_2) \gamma^{\xi} U(k_1)
, $$
$$
\mathcal{M}_{a.5} =  e  g^2    U(p_1) \gamma^{\xi} t^{\alpha}
U(k_1) {g^{\mu \xi} \delta^{\alpha \beta} \over (k_1-p_1)^2}
\bar{U}(p_2)\gamma^{\mu}    t^{\beta}  {\slashed{k_2}-\slashed{p}+m
\over(k_2-p)^2 - m^2}  \epsilon_\nu(p)  \gamma^{\nu} \bar{U}(k_2)  ,
    $$
   $$
         \mathcal{M}_{a.6} =    e  g^2    \bar{U}(p_2) t^{\alpha} \gamma^{\xi} U(p_1) {g^{\xi \nu} \delta^{\alpha \beta} \over (p_2+p_1)^2}  t^{\beta} U(k_1) \gamma^{\nu}  {\slashed{k_1}-\slashed{p}+m \over(k_1-p)^2 - m^2}        \gamma^{\mu}   \epsilon_\mu(p)    \bar{U}(k_2)  ,
  $$
$$
\mathcal{M}_{a.7} =   e  g^2    U(p_1) \gamma^{\nu} t^{\alpha}
\bar{U}(k_2) {g^{\nu \xi} \delta^{\alpha \beta} \over (k_2-p_1)^2}
\bar{U}(p_2)\gamma^{\xi} t^{\beta}  {\slashed{k_1}-\slashed{p}+m
\over(k_1-p)^2 - m^2}\epsilon_\nu(p)  \gamma^{\mu}  U(k_1)  ,
    $$
$$
\mathcal{M}_{a.8} = e  g^2   \bar{U}(p_2)\gamma^{\xi}
\epsilon_\xi(p){\slashed{p_2}+\slashed{p}+m \over(p_2+p)^2 -
m^2}U(k_1)\gamma^{\nu} t^{\alpha} {g^{\nu \nu} \delta^{\alpha \beta}
\over (k_2-p-p_2)^2}   t^{\beta}  U(p_1)   \gamma^{\mu}
\bar{U}(k_2)   ,
$$
6)\;  $q^*+\bar q^* \to \gamma+g+ g $ (NLO, figure \ref{fig10p}):
$$
\mathcal{M}_{b.1} =  e  g^2   U(k_1)   \gamma^{\mu}
\epsilon_\mu(p_2)  {\slashed{k_1}-\slashed{p_2}+m \over(k_1-p_2)^2 -
m^2} \gamma^{\nu}  \epsilon_\nu(p_1) {\slashed{k_2}-\slashed{p}+m
\over(k_2-p)^2 - m^2}    \epsilon_\xi(p) \gamma^{\xi}\bar{U}(k_2),
 $$
  $$
\mathcal{M}_{b.2} =   e  g^2   U(k_1)   \gamma^{\mu}
\epsilon_\mu(p_2)  {\slashed{k_1}-\slashed{p_2}+m \over(k_1-p_2)^2 -
m^2} \epsilon_\xi(p) \gamma^{\xi}{\slashed{k_2}-\slashed{p_1}+m
\over(k_2-p_1)^2 - m^2} \gamma^{\nu}  \epsilon_\nu(p_1)
\bar{U}(k_2),
    $$
$$
\mathcal{M}_{b.3} =  e  g^2   U(k_1)   \gamma^{\mu}
\epsilon_\mu(p_1)  {\slashed{k_1}-\slashed{p_1}+m \over(k_1-p_1)^2 -
m^2} \epsilon_\xi(p) \gamma^{\xi}{\slashed{k_2}-\slashed{p_2}+m
\over(k_2-p_2)^2 - m^2} \gamma^{\nu}  \epsilon_\nu(p_2 )
\bar{U}(k_2) , $$
$$
\mathcal{M}_{b.4} =  e  g^2  \bar{U}(k_2)    \gamma^{\mu}
\epsilon_\mu(p_2)  {\slashed{k_1}-\slashed{p_2}+m \over(k_1-p_2)^2 -
m^2} \gamma^{\nu}  \epsilon_\nu(p_1) {\slashed{k_2}-\slashed{p}+m
\over(k_2-p)^2 - m^2}    \epsilon_\xi(p) \gamma^{\xi} U(k_1), $$
$$
\mathcal{M}_{b.5} = e  g^2   U(k_1)   \gamma^{\nu}
\epsilon_\nu(p_1)  {\slashed{k_1}-\slashed{p_1}+m \over(k_1-p_1)^2 -
m^2} \epsilon_\mu(p_2) \gamma^{\mu}{\slashed{k_2}-\slashed{p}+m
\over(k_2-p)^2 - m^2} \gamma^{\xi}  \epsilon_\xi(p ) \bar{U}(k_2) ,
    $$
   $$
         \mathcal{M}_{b.6} =   e  g^2   U(k_1)   \gamma^{\xi}   \epsilon_\xi(p)  {\slashed{k_1}-\slashed{p}+m \over(k_1-p)^2 - m^2} \epsilon_\mu(p_2) \gamma^{\mu}{\slashed{k_2}-\slashed{p_2}+m \over(k_2-p_2)^2 - m^2} \gamma^{\xi}  \epsilon_\xi(p ) \bar{U}(k_2) ,
  $$
$$
\mathcal{M}_{b.7} =  e  g^2   U(k_1)   \gamma^{\nu}
\epsilon_\nu(p)C^{\nu \mu \rho}(p_1,p_2,(p_1+p_2))
{\epsilon_\mu(p_1)\epsilon_\rho(p_2)\over(p_1+p_2)^2 } f^{abc} t^a
{\slashed{k_2}-\slashed{p}+m \over(k_2-p)^2 - m^2}\gamma^{\xi}
\epsilon_\xi(p ) \bar{U}(k_2),
    $$
$$
\mathcal{M}_{b.8} = e  g^2   \bar{U}(k_2)   \gamma^{\nu}
\epsilon_\nu(p)C^{\nu \mu \rho}(p_1,p_2,(p_1+p_2))
{\epsilon_\mu(p_1)\epsilon_\rho(p_2)\over(p_1+p_2)^2 } f^{abc} t^a
{\slashed{k_2}-\slashed{p}+m \over(k_2-p)^2 - m^2}\gamma^{\xi}
\epsilon_\xi(p )   U(k_1) .
$$
where $t^a={\lambda^a\over 2}$, $\lambda^a$ are the Gell-Mann
matrices and $C^{\mu \nu \rho}(k_1,k_2,k_3)$ stand for the standard
three gluon coupling vertex:
$$g^{\mu \nu}(k_2-k_1)^\rho+g^{ \nu\rho}(k_3-k_2)^\mu+g^{
\rho\mu}(k_1-k_3)^\nu.
$$
$m$ and $e$ are the mass and the fractional electric charge of the
quark q. We use the algebraic manipulation system $\mathtt{FORM}$
\cite{FORM} to compute    the squared of the above amplitude in the
equations (\ref{eq251}), (\ref{eq252}) and (\ref{eq261}) with the
small x approximation, e.g., $U(p)\bar U(p)=x\slashed{p}$. The
polarization 4-vectors of the outgoing photon and gluon in the above
squared amplitude, e.g., $\epsilon^\xi(p)$, that satisfy the
co-variant equation \cite{amin} (note that they give only the
transverse part with respect to their momentums):
        \begin{equation}
\sum  \epsilon^{\mu}(p)  \epsilon^{*\nu}(p) =-g_{\perp}^{\mu \nu}.
    \label{eq002p}
    \end{equation}
On the other hand, $\epsilon^\mu(k_i)$ is the polarization vector of
the incoming off-shell gluons which should be modified with the
eikonal vertex (i.e the BFKL prescription, see the reference
\cite{Deak1}). One choice is to impose the so called non-sense
polarization conditions on $\epsilon^\mu(k_i)$ which is not
normalized to one \cite{Deak1,co} (and it will not be used in the
present work):
$$
    \epsilon^{\mu}(k_i) = {2 k^\mu_{i} \over \sqrt{s}}.
    $$
But in the case of $k_t$-factorization scheme and the off-shell
gluons, the better choice is $\epsilon^\mu(k_i)={k^\mu_{i,t} \over
|k_{i,t}|}$, which leads to the following identity and can be easily
implemented in our calculations \cite{Deak1,co,amin}:
         \begin{equation}
\sum \epsilon^{\mu}(k_i) \epsilon^{*\nu}(k_i)={{k_{i,t}^{\mu}
k_{i,t}^{\nu}}\over k_{i,t}^2}.
    \label{eq003}
    \end{equation}

\newpage
\begin{figure}[H]
\centering
\includegraphics[scale=0.3]{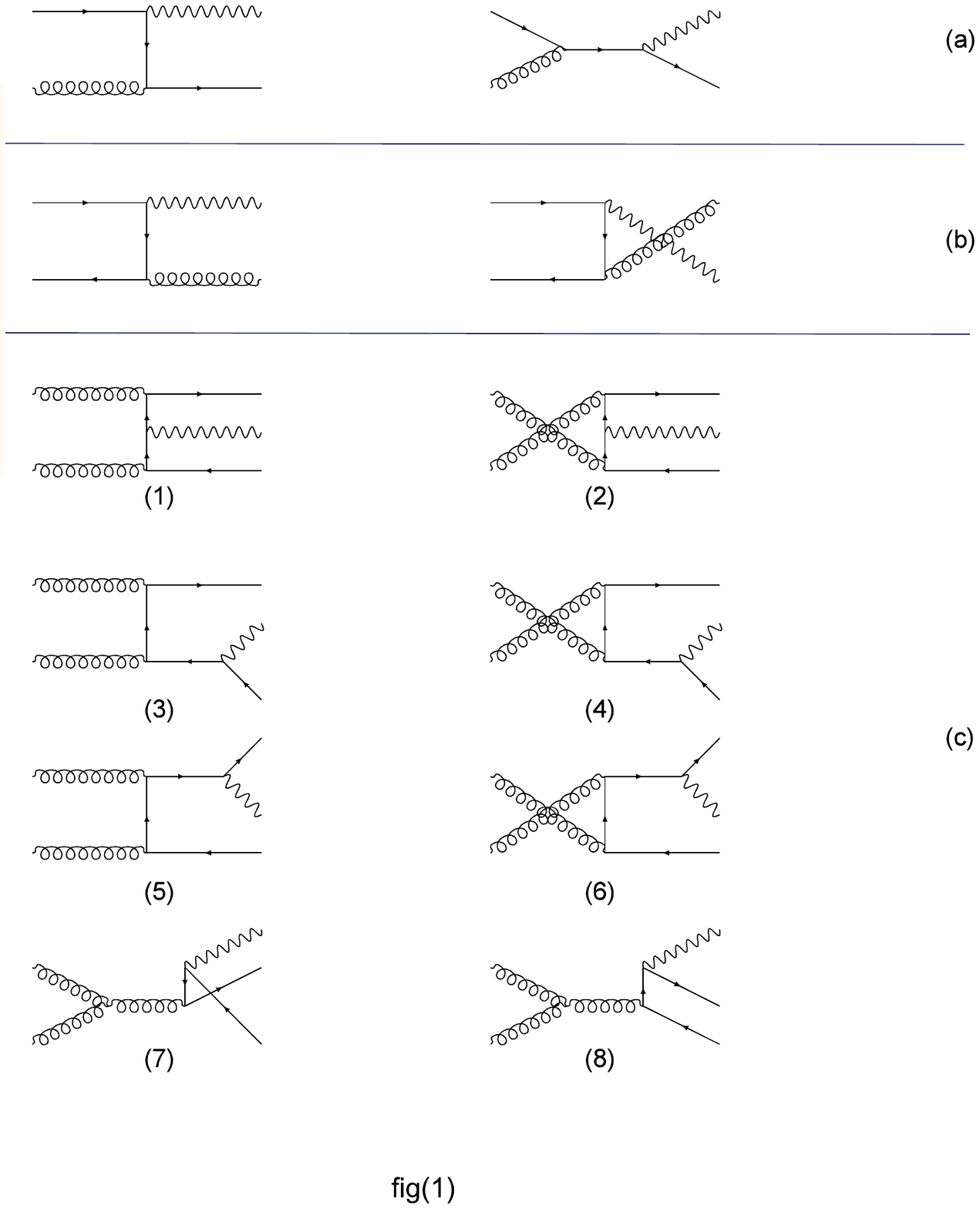}
\caption{The different Feynman diagrams of prompt single-photon
production in the lowest order in each sub-process, i.e., (a)
$q(\bar q)+g \to\gamma+ q (\bar q)$ , (b) $q+\bar q\to \gamma+ g$
and (c) $g+g\to\gamma+ q + \bar q$.} \label{fig0}
\end{figure}
\begin{figure}[H]
\centering
\includegraphics[scale=0.3]{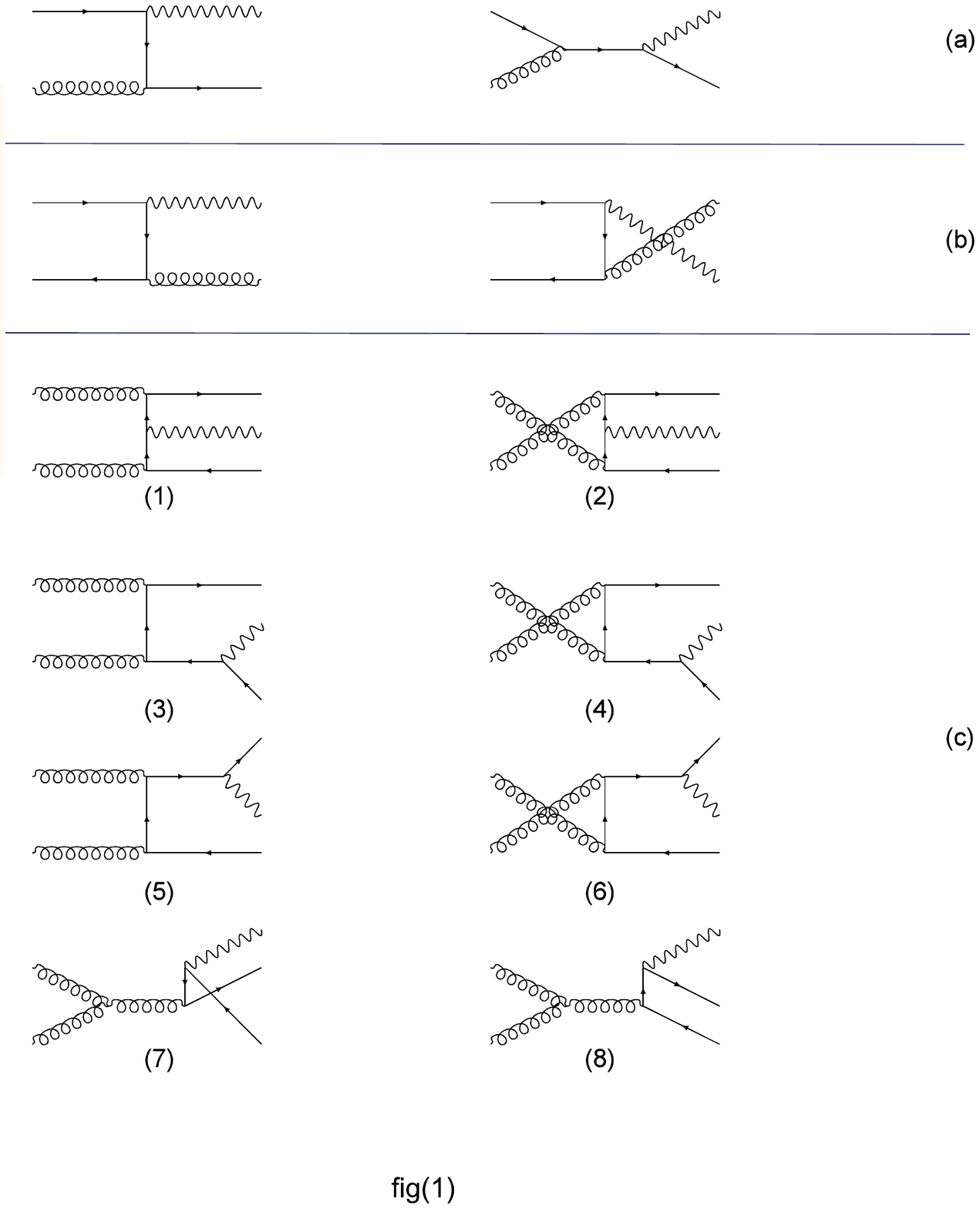}
\caption{The double-differential cross-section for the production of
a single-photon as a function of the transverse momentum of the
resulting photon. The corresponding numerical calculations are
carried out within the given rapidity region, utilizing the UPDF of
KMR, LO and NLO MRW for $E_{CM} = 1.96$ TeV. The results are shown
in the panels (a), (b) and (c), respectively. These panels also
outline the contributions of the involving partonic subprocesses.
The uncertainty regions are designated via manipulating of the
hard-scale of the processes by a factor of $2$. The panel (d)
presents a comparison between these results against each other and
those of the experimental data of the $D0$ collaboration at each
bin, \cite{D02005} as well as those of JETPHOX
\cite{jetphox,D02005}. In the bottom of this panel the ratio of
differential cross-sections ($\cal R$) to that of experimental data
is also presented (red-circle, green-triangle and blue-square are
for KMR, LO MRW and NLO-MRW, respectively ).} \label{fig1}
\end{figure}

\begin{figure}[ht]
\centering
\includegraphics[scale=0.3]{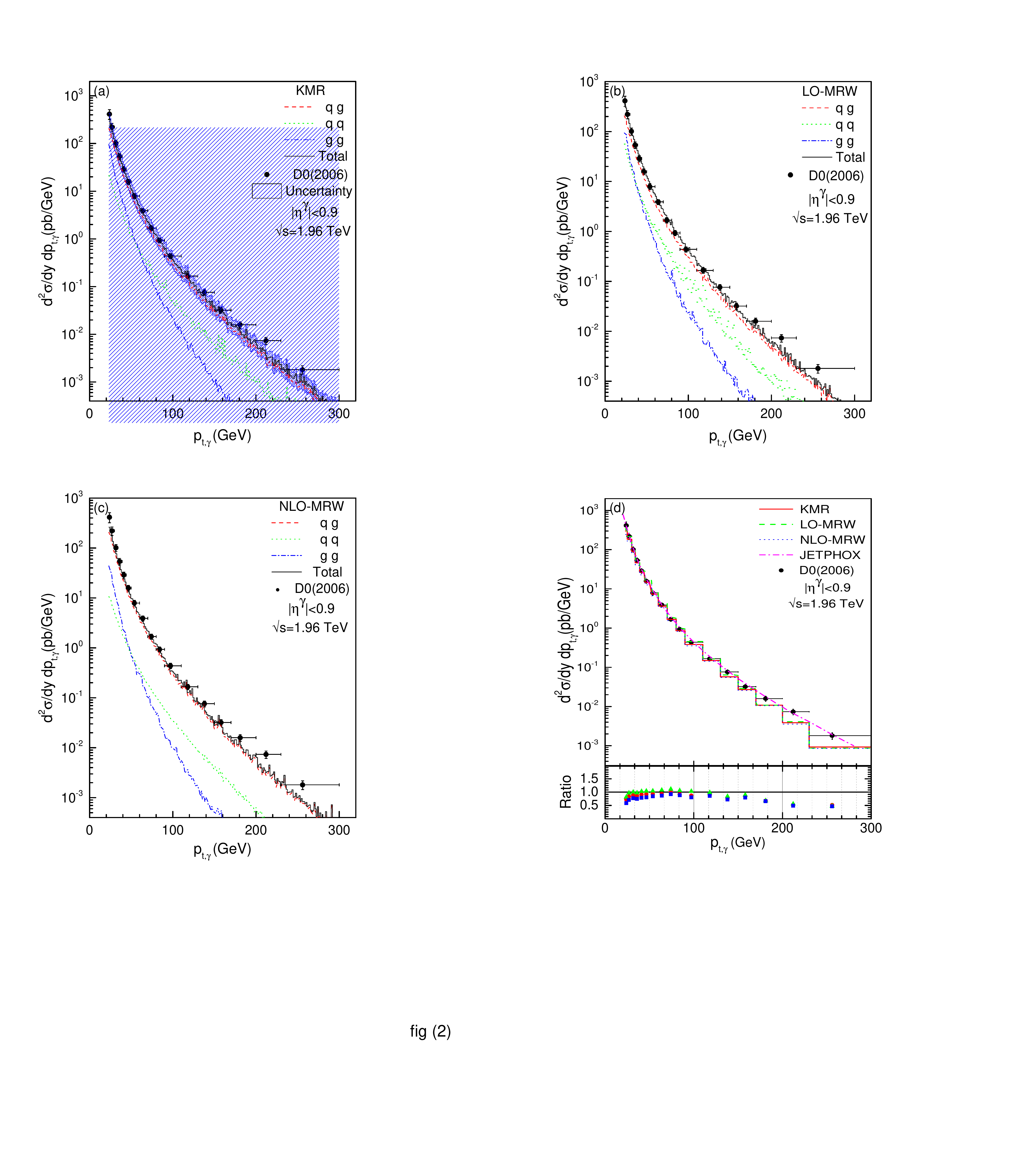}
\caption{The double-differential cross-section for the production of
a single-photon as a function of the transverse momentum of the
resulting photon. The notion of the figure is as in the figure
\ref{fig1} with this difference that the data are selected from the
CDF collaboration \cite{CDF2017} and also in the panel (d) the
comparison is made with the results of  SHERPA , PYTHIA and MCFM
 \cite{sherpa,pythia, mcfm,CDF2017}.} \label{fig2}
\end{figure}

\begin{figure}[ht]
\centering
\includegraphics[scale=0.3]{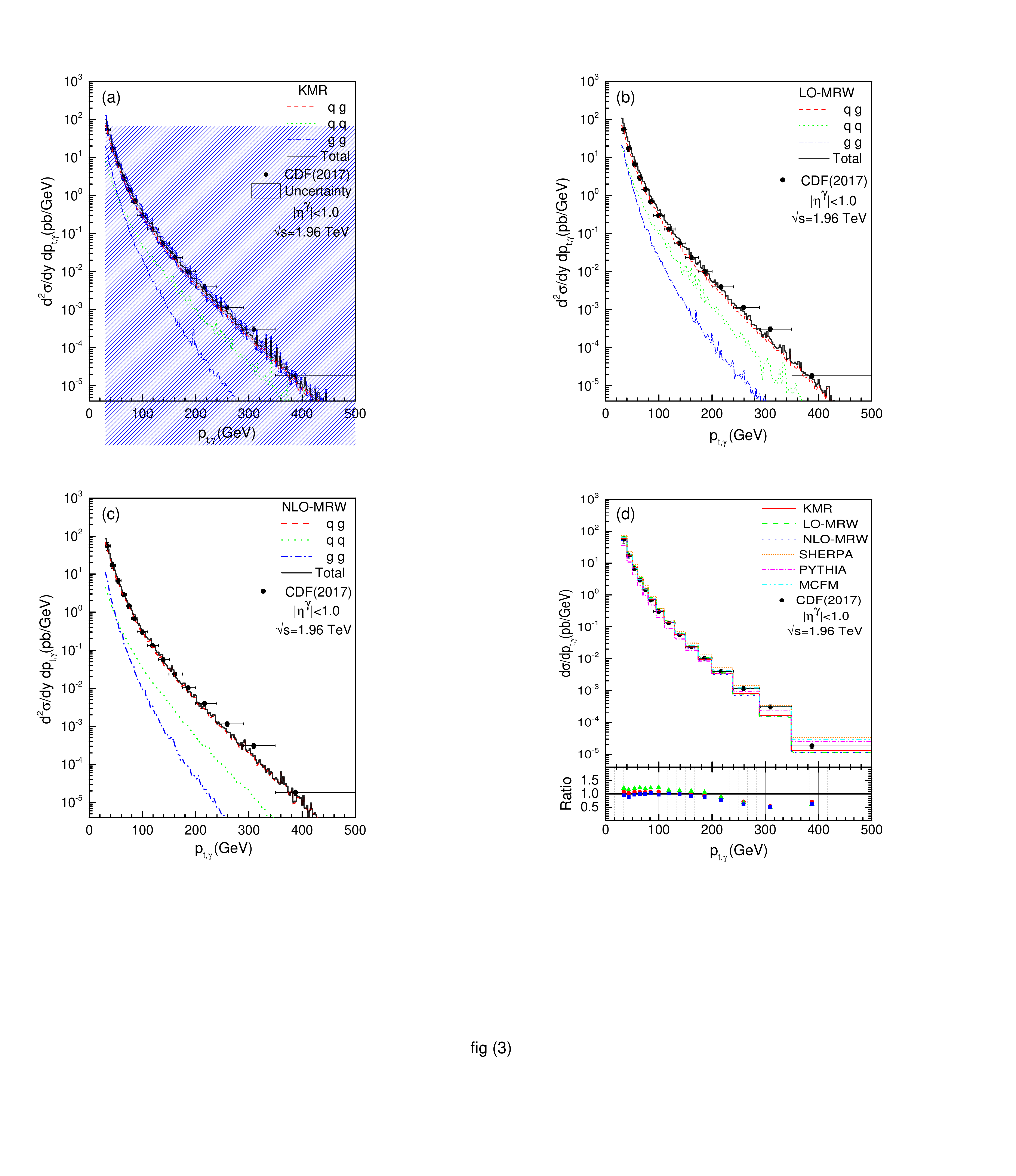}
\caption{The differential cross-section for the production of a
single-photon as a function of the transverse momentum of the
resulting photon. The corresponding numerical calculations are
carried out within the given rapidity boundaries (see the legends of
the plots), utilizing the UPDF of KMR, LO and NLO MRW for $E_{CM} =
13$ TeV. The results are shown in the panels (a), (b) and (c)
respectively. These panels also outline the contributions of the
involving partonic sub-processes. The uncertainty regions are
designated via manipulating of the hard-scale of the processes by a
factor of $2$. The panel (d) presents a comparison between these
results against each other and those of the experimental data of the
ATLAS collaboration  at each bin, as well as SHERPA and PYTHIA,
\cite{sherpa, pythia,Atlas2017}. In the bottom of this panel the
ratio of differential cross-sections ($\cal R$) to that of
experimental data is also presented (red-circle, green-triangle and
blue-square are for KMR, LO MRW and NLO MRW, respectively ).}
\label{fig3}
\end{figure}

\begin{figure}[ht]
\centering
\includegraphics[scale=0.3]{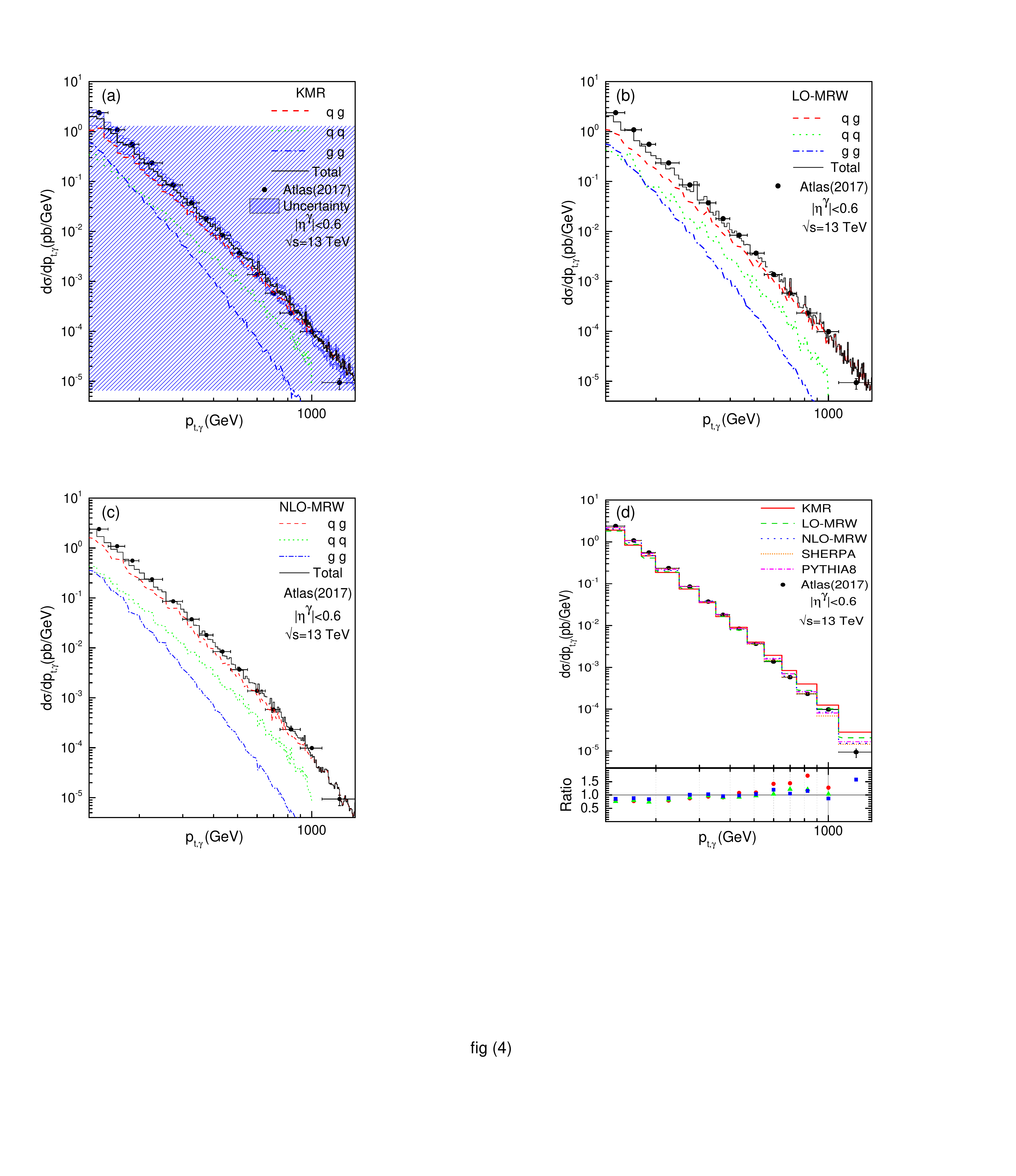}
\caption{The differential cross-section for the production of a
single-photon as a function of the transverse momentum of the
resulting photon. The notion of the figure is as in the figure
\ref{fig3}.} \label{fig4}
\end{figure}

\begin{figure}[ht]
\centering
\includegraphics[scale=0.3]{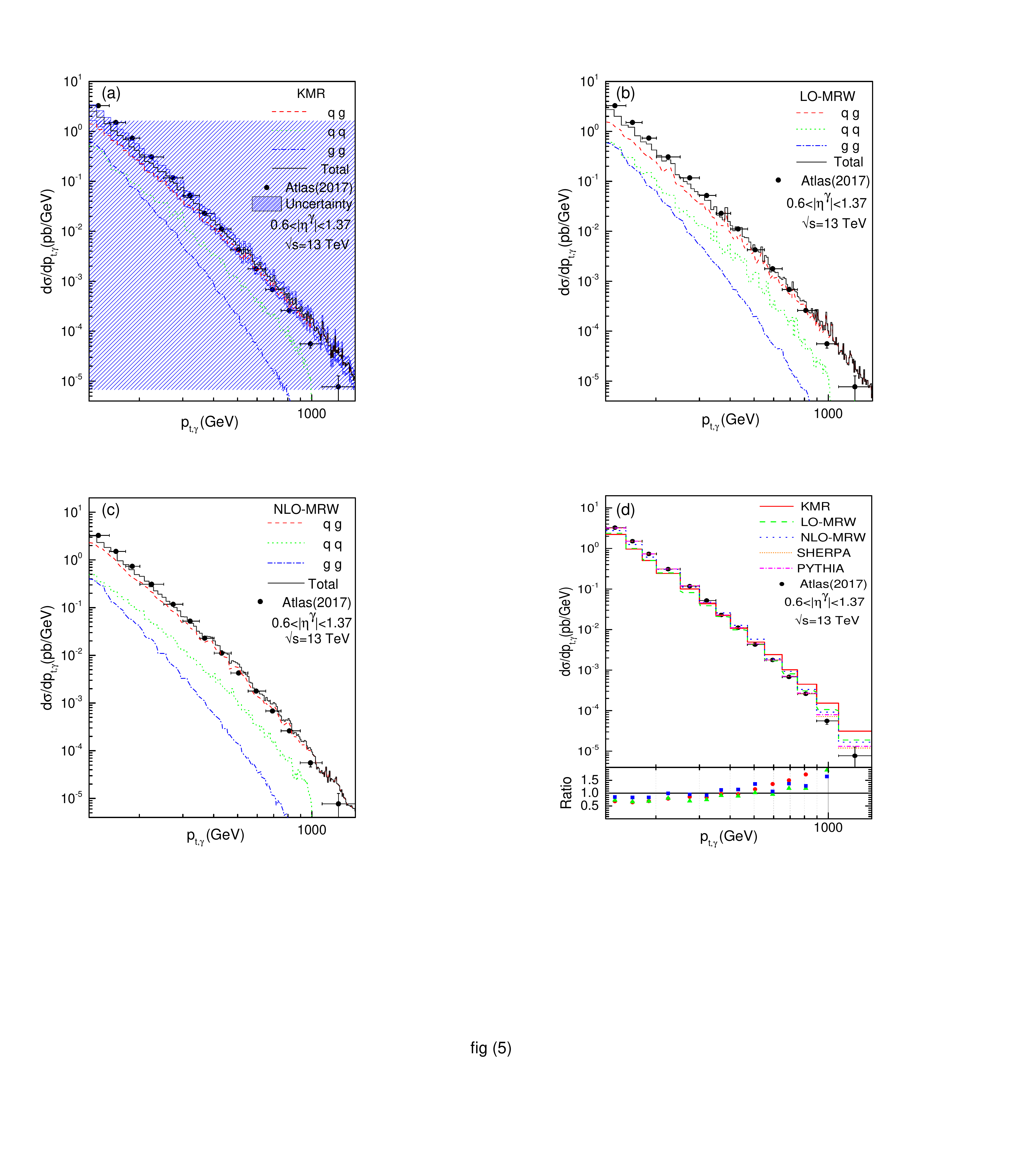}
\caption{The differential cross-section for the production of a
single-photon as a function of the transverse momentum of the
resulting photon. The notion of the figure is as the figure
\ref{fig3}.} \label{fig5}
\end{figure}

\begin{figure}[ht]
\centering
\includegraphics[scale=0.3]{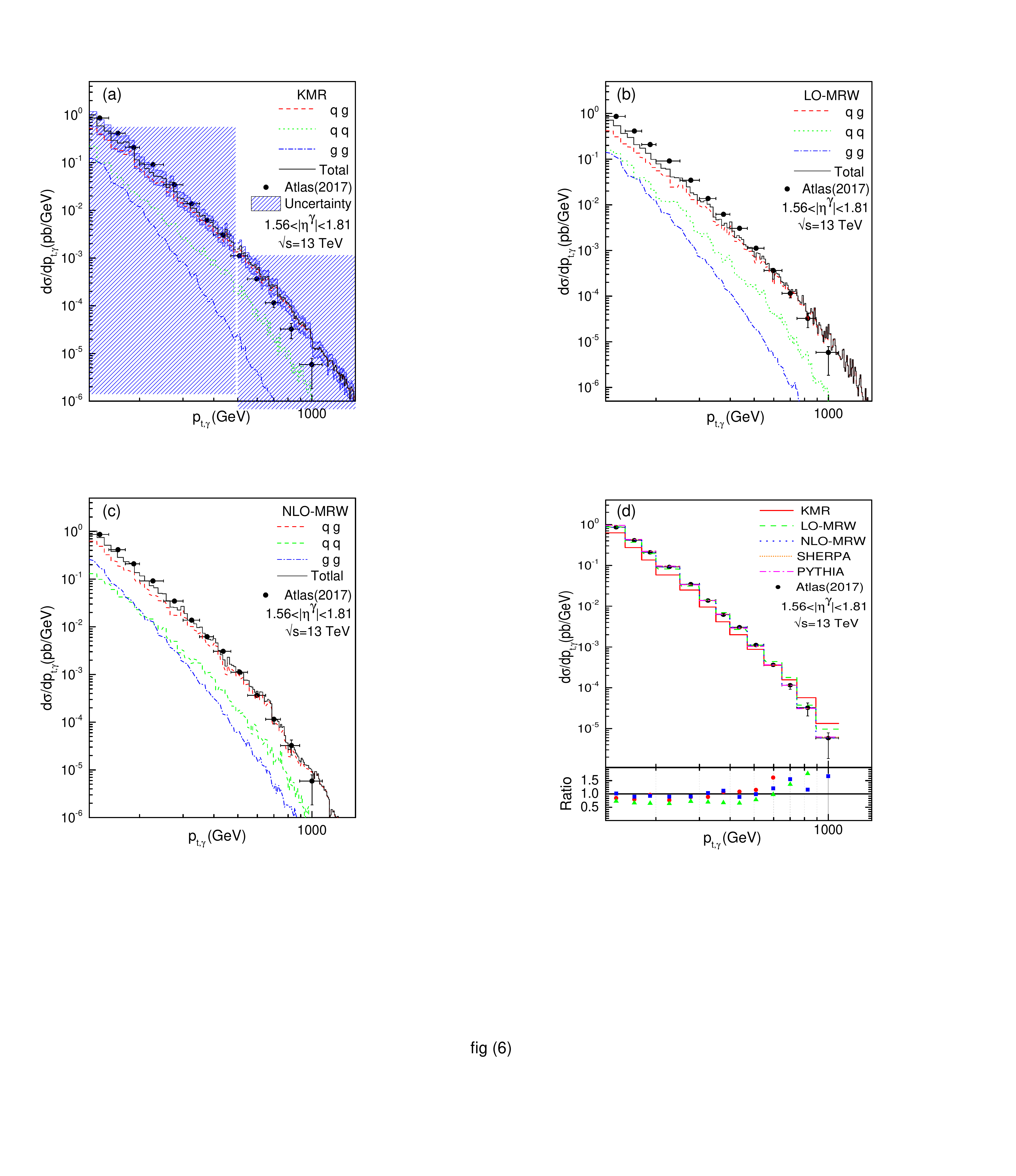}
\caption{The differential cross-section for the production of a
single-photon as a function of the transverse momentum of the
resulting photon. The notion of the figure is as in the figure
\ref{fig3}.} \label{fig6}
\end{figure}

\begin{figure}[ht]
\centering
\includegraphics[scale=0.3]{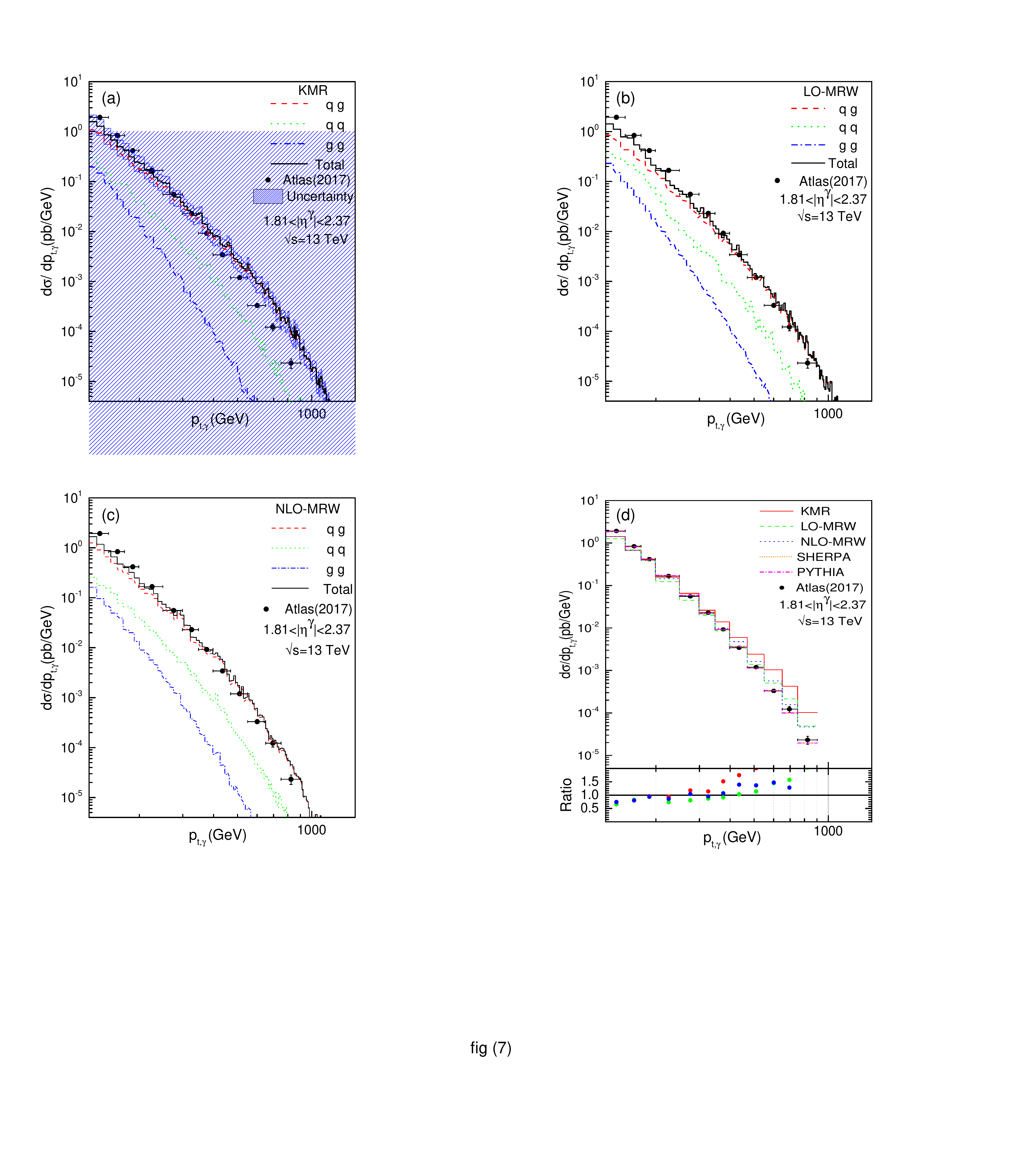}
\caption{The double-differential cross-section for the production of
a single-photon as a function of the transverse momentum of the
resulting photon. The corresponding numerical calculations are
carried out within the given rapidity region, utilizing the UPDF of
KMR, LO and NLO MRW for $E_{CM} = 7$ TeV. The results are shown in
the panels (a), (b) and (c), respectively. These panels also outline
the contributions of the involving partonic sub-processes. The
uncertainty regions are designated via manipulating the hard-scale
of the processes by factor of $2$. The panel (d) presents a
comparison between these results against each other and those of the
experimental data of the CMS collaboration at each bin, \cite{cms}
as well as JETPHOX  \cite{jetphox,cms}.  In the bottom of this panel
the ratio of differential cross-sections ($\cal R$) to that of
experimental data is also presented (red-circle, green-triangle and
blue-square are for KMR, LO MRW and NLO MRW, respectively ).}
\label{fig7}
\end{figure}
\begin{figure}[ht]
\centering
\includegraphics[scale=0.3]{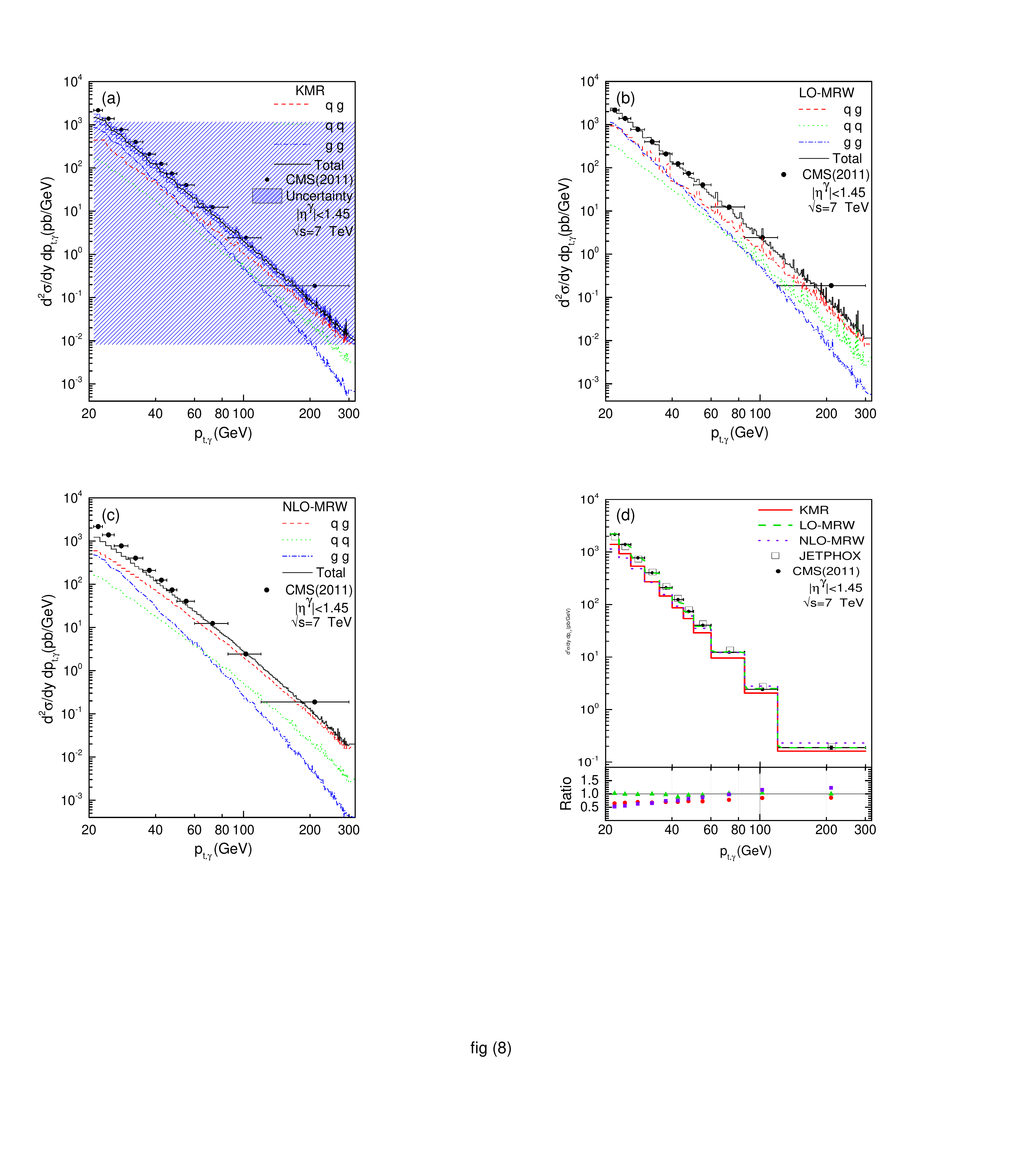}
\caption{The different Feynman diagrams of prompt single-photon
production in the NLO order for the $q(\bar q)+g \to \gamma+q
(\bar q)+g $ sub-process.}
 \label{fig9}
 \end{figure}
\begin{figure}[ht]
\centering
\includegraphics[scale=0.3]{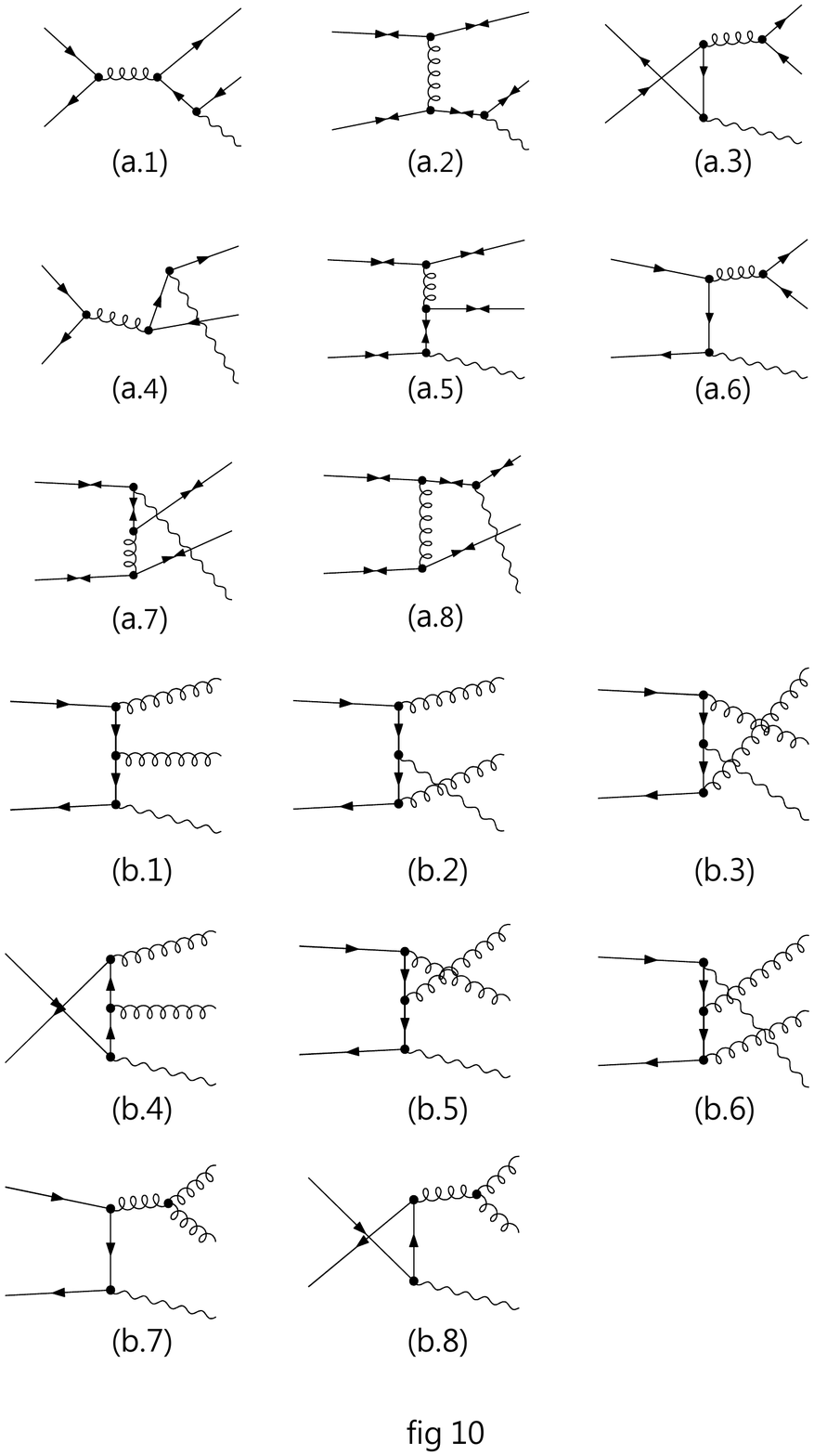}
\caption{The different Feynman diagrams of prompt single-photon
production in the NLO level for the (a) $q(\bar q)+q(\bar q)  \to \gamma+q(\bar q)+
q(\bar q) $  and (b) $q+\bar q \to \gamma+g+ g $  sub-processes.}
 \label{fig10p}
 \end{figure}
\begin{figure}[ht]
\centering
\includegraphics[scale=0.3]{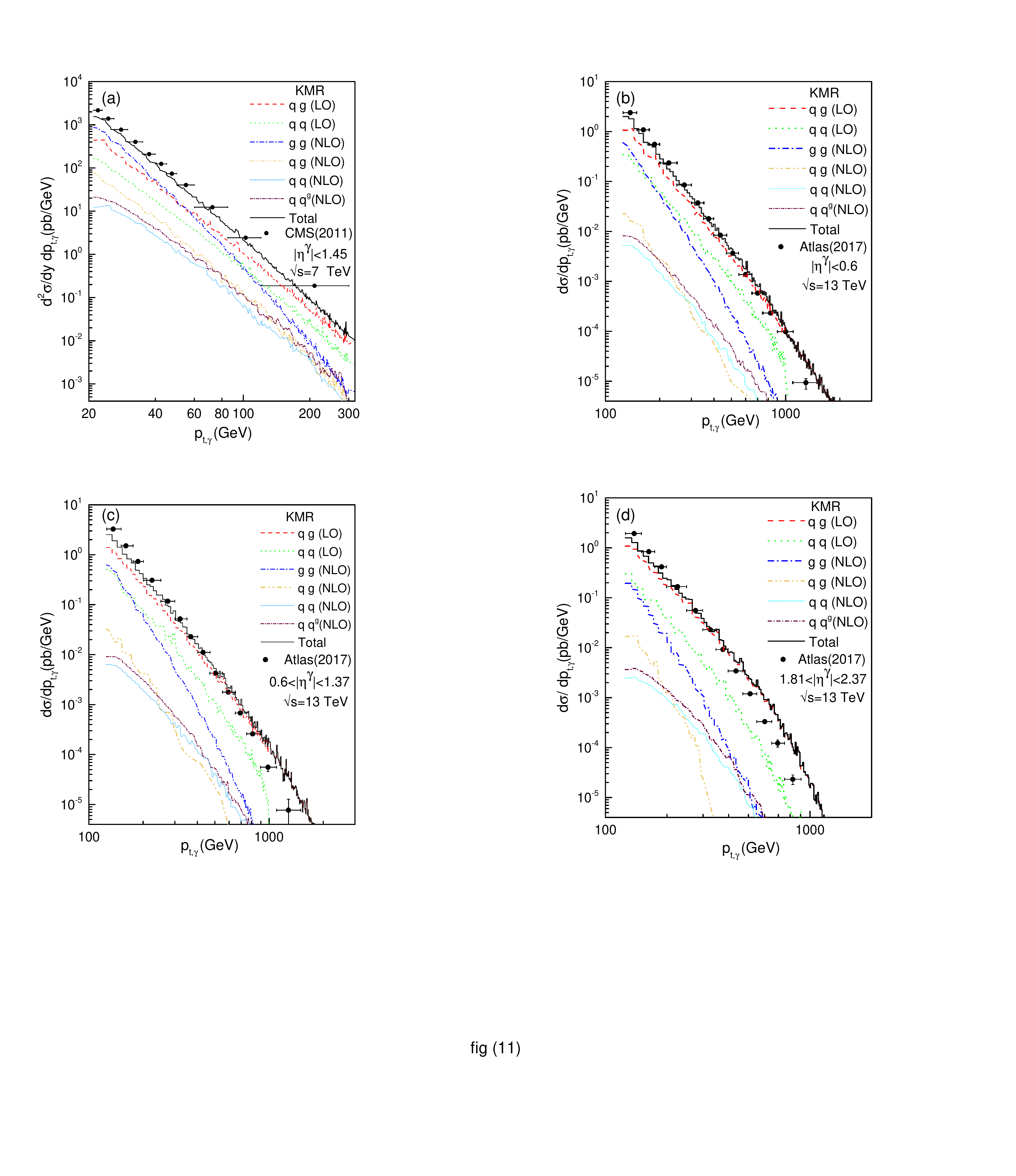}
\caption{The differential cross-sections for the production of a
single-photon as a function of the transverse momentum of the
resulting photon. The corresponding numerical calculations are
carried out within the given rapidity boundaries (see the legends of
the plots), utilizing the UPDF of KMR. The results of $q(\bar
q)+g \to \gamma+q (\bar q)+g $, $q(\bar q)+q(\bar q)  \to \gamma+q(\bar q)+
q(\bar q) $ and $q+\bar q \to \gamma+g+ g $ sub-processes in the
NLO level are shown in the legend of  panels (a), (b), (c) and (d)
by  $q g$, $q q$ and $q q^g$, respectively.} \label{fig10}
\end{figure}
\begin{figure}[ht]
\centering
\includegraphics[scale=0.3]{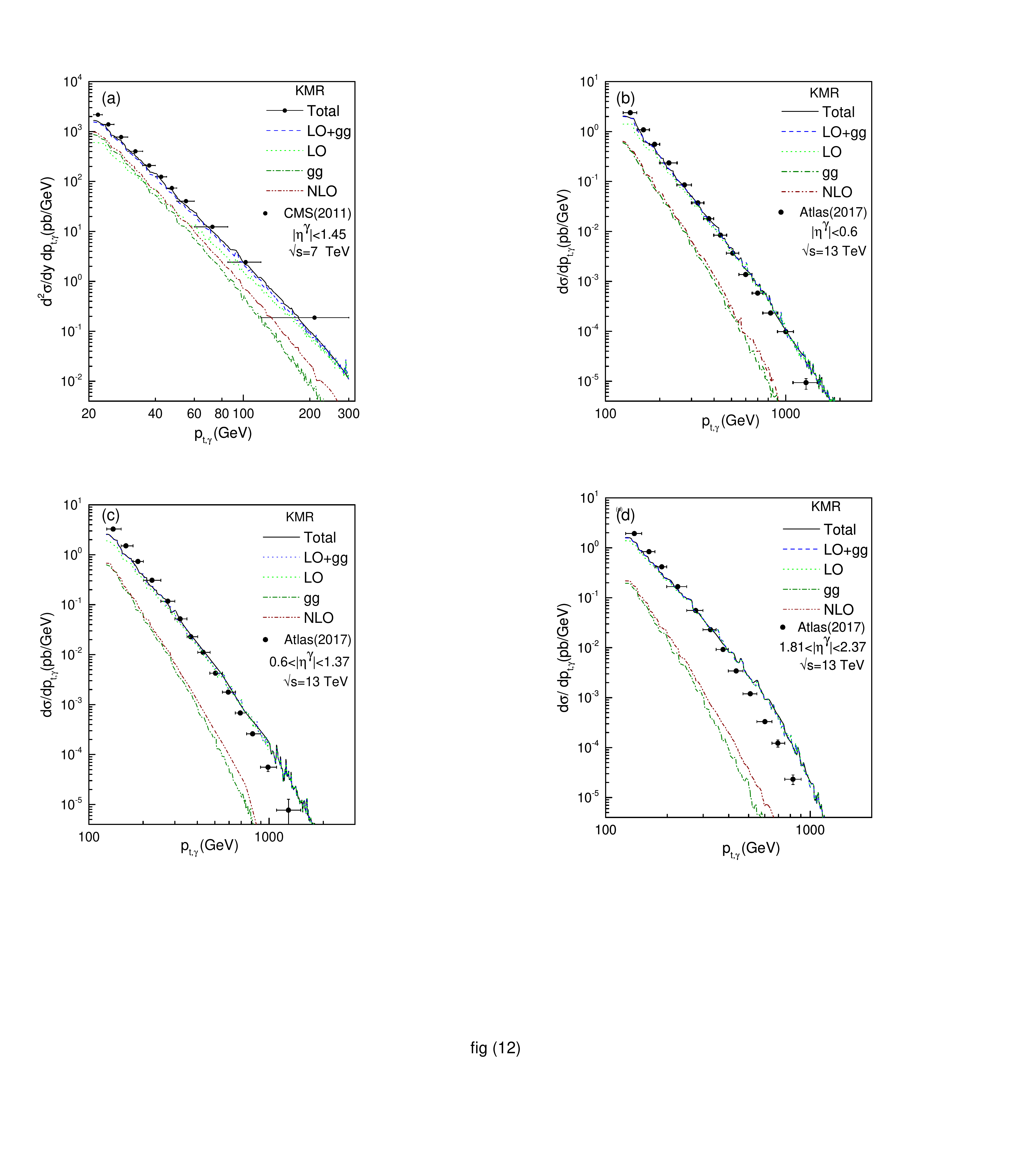}
\caption{As the figure \ref{fig10} but for different contributions of the LO and NLO levels to the differential cross-sections, see the legend of the figure and the explanations in text.}
 \label{fig12}
 \end{figure}

\newpage
\end{document}